\documentclass{JHEP3}
\usepackage{epsfig,amssymb,amsmath,latexsym}

\newcommand{\be}{\begin{equation}}
\newcommand{\ee}{\end{equation}}
\newcommand{\bea}{\begin{eqnarray}}
\newcommand{\ena}{\end{eqnarray}}
\newcommand{\ba}{\begin{eqnarray}}
\newcommand{\ea}{\end{eqnarray}}

\newcommand{\nb}{\nonumber}

\renewcommand{\H}{{\mathcal H}}
\newcommand{\K}{{\mathcal K}}
\newcommand{\B}{{\mathcal B}}
\newcommand{\A}{{\mathcal A}}
\newcommand{\M}{{\mathcal M}}
\newcommand{\LL}{{\mathcal L}}

\newcommand{\pd}{\partial}
\newcommand{\di}{\mathrm d}
\newcommand{\de}{\partial}

\renewcommand\S{\Sigma}

\renewcommand\a{\alpha}
\renewcommand\b{\beta}
\newcommand\m{\mu}

\newcommand\n{\nu}

\newcommand\s{\sigma}
\renewcommand\t{\ensuremath{\tau}}
\renewcommand\l{\ensuremath{\lambda}}

\renewcommand\L{\ensuremath{\Lambda}}

\renewcommand\underline{}
\renewcommand\mathbf{}

\makeatletter
\def\ps@mine{%
    \def\@oddfoot{\hfil\thepage\hfil}\let\@evenfoot\@oddfoot
    \def\@evenhead{\textsl{\jobname}\dotfill\textsl{\today}}%
    \let\@oddhead\@evenhead%
    \let\@mkboth\@gobbletwo
    \let\sectionmark\@gobble
    \let\subsectionmark\@gobble
    }

\title{
  Lorentz Breaking Massive Gravity in Curved Space
  }

\date{\today}

\author{
D.Blas,\!$^a$ D. Comelli,\!$^b$ F. Nesti,\!$^c$ L. Pilo$^c$\\
 \llap{$^a$}FSB/ITP/LPPC,
 \'Ecole Polytechnique F\'ed\'erale de Lausanne,\\ CH-1015, Lausanne, Switzerland\\
  \llap{$^b$}INFN, Sezione di Ferrara,  I-35131 Ferrara, Italy\\
  \llap{$^c$}Dipartimento di Fisica, Universit\`a di L'Aquila, I-67010 L'Aquila, and\\
  INFN, Laboratori Nazionali del Gran Sasso, I-67010 Assergi,
Italy}

\abstract{%
  A systematic study of the different phases of Lorentz-breaking
  massive gravity in a curved background is performed. For tensor and
  vector modes, the analysis is very close to that of Minkowski space.
  The most interesting results are in the scalar sector where,
  generically, there are two propagating degrees of freedom (DOF).
  While in maximally symmetric spaces ghost-like instabilities are
 inevitable, they can be avoided in a FRW background. The phases
  with less than two DOF in the scalar sector are also studied.
  Curvature allows an interesting interplay with the mass parameters;
  in particular, we have extended the Higuchi bound of dS to FRW and
  Lorentz breaking masses. As in dS, when the bound is saturated there
  is no propagating DOF in the scalar sector. In a number of phases the smallness of the
  kinetic terms gives rise to strongly coupled scalar modes at low
  energies. Finally, we have computed the gravitational potentials for
  point-like sources. In the general case we recover the GR
  predictions at small distances, whereas the modifications appear at
  distances of the order of the characteristic mass scale. In contrast
  with Minkowski space, these corrections may not spoil the linear
  approximation at large distances.}

\begin{document}

%%%%%%%%%%%%%%%%%%%%%%%%%%%%%%%%%%%%%%%%%%%%%%%%%%%%%%%%%%%%%%%%%
\section{Introduction}
%%%%%%%%%%%%%%%%%%%%%%%%%%%%%%%%%%%%%%%%%%%%%%%%%%%%%%%%%%%%%%%%%

Massive gravity has recently received a lot of attention mainly due to
its relation to large distance modifications of the gravitational
force (for recent reviews, see
e.g.~\cite{Rubakov:2008nh,Blas:2008uz}). Even if the addition of a
Lorentz-invariant mass term to the standard action for the graviton in
a flat background accomplishes the desired modification, it also
implies the appearance of new problems, such as the vDVZ discontinuity
and the strong coupling of the scalar mode of the massive
graviton \cite{vDVZ,ArkaniHamed:2002sp,Deffayet:2001uk,Dvali:2006su}.
It was realized in \cite{Rubakov:2004eb} (see also
\cite{Dubovsky:2004sg}) that some of these problems may be softened if
the mass term breaks the Lorentz invariance to rotational invariance.
 
\pagebreak[3] 

It is also known that some of the features of Lorentz breaking massive
gravity are peculiar to Minkowski space and do not hold in other
backgrounds \cite{vDVZAdS} (see also \cite{Creminelli:2005qk,Nair:2008yh}). In this
note, we will study the behavior of the gravitational perturbations in a
curved background with a mass term breaking linearized general covariance.

A caveat to this restriction has to do with the choice of action for
the graviton fluctuations in curved spacetime. Normally, one considers
the perturbations of the General Relativity (GR) action to second order around a background
which solves the equations of motion (EOM). This ensures gauge
invariance under linearized diffeomorphisms (diff) at the quadratic
level. In this work, we modify GR by adding a mass term which breaks
explicitly gauge invariance, but other diff-breaking corrections
are possible. In particular, also the kinetic term may be modified
once one relaxes the constraint of general
covariance.\footnote{Recently, there has been some interest in the
  modifying the kinetic structure of GR as a way to improve its UV
  behavior \cite{Horava:2009uw}.} The motivation for considering only
mass terms is that we want to focus on large distance (infrared)
modifications of gravity. Besides, there are known physical examples
generating this kind of mass terms for gravitational perturbations.

A first example is a model where the matter sector includes four
scalar fields that condense, breaking spontaneously 
the symmetry of the background metric~\cite{Dubovsky:2004ud} (see
also~\cite{ArkaniHamed:2003uy} for related previous work
and~\cite{Bebronne:2007qh} for some cosmological implications). In the
gauge where those scalar modes are frozen (unitary gauge), the
spectrum of the perturbations reduces to the gravitational modes with
a mass term that violates the symmetry of the background. In this
sense, the scalars are the Goldstones modes of the broken diff
invariance. Another interesting example is bigravity, where a second
rank-2 tensor interacts with the metric
$g_{\m\n}$~\cite{Damour:2002ws}. In this case, there are exact flat
backgrounds where the metrics do not share the whole group of
invariance, but preserve a common $SO(3)$.
The spectrum of fluctuations around these backgrounds includes a
Lorentz breaking massive graviton which is a combination of both
metrics~\cite{Blas:2007zz,Berezhiani:2007zf} (see
also~\cite{Berezhiani:2009kx} for some phenomenology and
\cite{Berezhiani:2008nr,Deffayet:2008zz} for spherically symmetric
solutions).

Inspired by the previous models\footnote{Other related models include
  theories with extra dimensions \cite{Koroteev:2009xd} and theories with
  condensing vector fields~\cite{Libanov:2005vu}.} we will consider
the presence in the action of a generic mass term (function of the
metric) which breaks general covariance. This term both allows for FRW
backgrounds (see e.g.~\cite{Dubovsky:2005dw,Grisa:2008um}) and 
generates the LB mass terms for the gravitational perturbations. Thus,
the analysis of just the gravitational degrees of freedom is
consistent in this setup where diff-invariance is broken, while in a
diff-invariant context this is possible only in a de Sitter (dS)
space.\footnote{We leave the analysis of gravitational perturbations
  coupled to additional fields in a FRW background for a forthcoming
  publication \cite{BCNP}.}
 
The paper is organized as follows: In section~\ref{sec:action} we introduce
our notations and the setup of our investigations. Then we analyze the
perturbations with LB terms in curved backgrounds for the tensor
(section~\ref{sec:tensors}), vector (section~\ref{sec:vectors}) and scalar
modes (section~\ref{sec:Scalars} and appendix~\ref{app1}). In
section~\ref{sec:vDVZ} we study the generalized Newton-like potentials and their
deviations from GR. We present the conclusions in
section~\ref{sec:conclusions}.

%%%%%%%%%%%%%%%%%%%%%%%%%%%%%%%%%%%%%%%%%%%%%%%%%%%%%%%
\section{Action, background and perturbations}
%%%%%%%%%%%%%%%%%%%%%%%%%%%%%%%%%%%%%%%%%%%%%%%%%%%%%%%
\label{sec:action}

Our starting point is the Einstein-Hilbert (EH)
Lagrangian with the addition of mass terms for the gravitational
perturbations, breaking general covariance. In the flat limit these
terms also break Lorentz invariance, and we will refer to them as
Lorentz-breaking (LB) terms.

This setup describes, at quadratic level, infrared modifications of
gravity where only gravitational degrees of freedom are present. At
full nonlinear level these deformations of the EH theory may be
parametrized by adding to the lagrangian a \emph{nonderivative}
function of the metric components, breaking general covariance:
\be
\label{ugac}
S=\int \!\di^4x\,\sqrt{-g}\, M_{P}^2\big[R - 2F(g_{\mu\nu})\big] \,.
\ee
General covariance can be
restored~\cite{ArkaniHamed:2002sp,Dubovsky:2004sg} by introducing
extra (St\"uckelberg) fields,\footnote{This implies the addition of
  (at most) four scalar fields, but invariant actions can also be
  found by adding vector or tensor fields.} which in the equivalent of
the unitary gauge yield the form (\ref{ugac}). 

It is clear that the term $F$ will contribute to the background EOM,
and exact solutions are known for certain $F$ functions. For example,
when $F=2\l=$ constant, the homogeneous and isotropic (FRW) background
will be maximally symmetric and the theory will be gauge (diff)
invariant. For $F\neq$ constant, FRW solutions can be found, which can
modify the standard cosmological solutions of GR. For certain classes
of $F$, solutions that exhibit late-time cosmic acceleration were
studied in~\cite{Dubovsky:2004ud} (see also \cite{Grisa:2008um}).

Accordingly, we assume that the dynamics of modified gravity admits a
spatially flat isotropic and homogeneous background (FRW henceforth)
\be
\bar g_{\m\n}=a(\eta)^2\eta_{\m\n}\qquad
\text{with}\ \ \eta_{\m\n} =\mathrm{diag}(-1,1,1,1)\,,
\ee
where $\eta$ is the conformal time. We will use
$\H(\eta)=a'/a$ and $H(\eta)=a'/a^2$, where ${}'$ is
the derivative with respect to $\eta$ (therefore $a H'=\H'-\H^2$).

We define the metric perturbations as
\be
g_{\m\n}=a^2\left(\eta_{\m\n}+h_{\m\n}\right).
\label{pert}
\ee
The second-order expansion of $S$ can then be written as:
\be
S=S_{GR}^{(2)}+ S_{LB}^{(2)}\,,
\label{eq:2nd}
\ee
where\footnote{We stress that in this expression, $H$ depends only on the background.}
\be
\label{GR}
 S_{GR}=\int \!\di^4x\,\sqrt{-g}\,  M_{P}^2(\,R-6H^2)
\ee
and the term $F$ gives rise also to LB masses for $h$. Assuming
rotations are preserved, these can be parametrized as
\be
S_{LB}^{(2)}=
\frac{M_{P}^2}{4}\int \!\di^4x\, \sqrt{-\bar g}\Big[m_0^2
\,h_{00}^2 + 2 m_1^2\, h_{0i}^2 - (m_2^2-4H'a^{-1})h_{ij}^2 + (m_3^2-2H'a^{-1})h_{ii}^2
- 2 m_4^2\, h_{00}h_{ii} \Big].
\label{LBmass}
\ee
Here spatial indices are contracted with $\delta_{ij}$, and $m_i\equiv
m_i(\eta)$ represent effective time dependent masses. The terms
proportional to $H'$ in (\ref{LBmass}) are conveniently chosen to
cancel similar contributions coming from the expansion of (\ref{GR})
in backgrounds different from dS. Notice that the parametrization in
(\ref{LBmass}) is completely general as the mass parameters are
arbitrary functions of the conformal time.

Diff gauge invariance is restored taking the limit $m_i=0$, $H'=0$ and
it corresponds to the case $F=\lambda=$constant. On the other hand,
for FRW background and vanishing masses the action is invariant only
under longitudinal spatial diffs. This is a consequence of the fact
that a generic FRW background is never a consistent background for GR
without matter. A non maximally symmetric background breaks the time
diffs, and accordingly in the limit of vanishing masses one recovers
the invariance under spatial diffs.
 
In the Lorentz-invariant case the masses can be expressed in terms of 
two parameters $\alpha$, $\beta$
\be
\label{eq:PFlimit}
m_0^2 = \alpha + \beta \, , \quad m_1^2 = -\alpha \, , \quad m_2^2 - 4 a^{-1} H' =
- \alpha \, , \quad m_4^2 = \beta  \, , \quad  m_3^2 - 2 a^{-1} H' = \beta \, ,
\ee
and the mass term (\ref{LBmass}) can be written in terms of contractions of
$h_{\m\n}$ with $\bar g^{\m\n}$.
The Fierz-Pauli (FP) choice, free of ghosts in flat space, corresponds
to $\alpha + \beta=0$. In curved space also the ``non-Fierz-Pauli''
case $\alpha+\beta\neq0$ can be free of ghosts (see
section~\ref{sec:Scalars}).

The setup introduced here is suitable to describe a rather general class of
massive gravity theories, and exhibits a rich set of phases, depending on
the masses $m_i$ and $H(\eta)$. In general we can have the following
scenarios:
\begin{itemize}
\item
  The $F$ term in the action (\ref{ugac}) does not affect neither the
  background (dS limit) nor the propagation of the perturbations. This is
 the case for $H'=0$ and $m_i= 0$ and is realized, e.g., when
  $F=\lambda$.

\item Only the perturbations are modified. This corresponds to $H'= 0$ and
  $m_i\neq 0$. Writing $F(g_{\mu\nu})=\lambda+f(g_{\mu\nu})$, this happens
  when the scale of $\lambda$ is much larger than the scale related to
  $f$.

\item Only the background is modified. This happens for $H'\neq 0$ and
  $m_i= 0$. As we will see, it can also be realized in less trivial situations.

\item Finally, in general both the
  background and the perturbations are modified by $F$.
\end{itemize}
\smallskip
In order to study the dynamics of the perturbations, it is convenient to
decompose the metric fluctuations as irreducible representations of the
rotation group:\footnote{Here, we follow the notation of
  \cite{Rubakov:2004eb}. }
\be
\begin{split}
\label{decomp}
& {h}_{00} = \psi \, ,\\
& {h}_{0 i} = {u}_i + \de_i v  \, , \qquad\qquad \qquad  \qquad \qquad \qquad \qquad
\de_i  {u}_i= 0,  \\
& {h}_{ij} = {\chi}_{ij} + \de_i s_j + \de_j s_i + \de_i \de_j \sigma +
\delta_{ij} \, \tau, \qquad
\de_i  s_i= \de_j {\chi}_{ij} = \delta_{ij} {\chi}_{ij} = 0 \,.
\end{split}
\ee
{}From those fields one can define two scalar and one vector gauge invariant quantities
\be
\begin{split}
&\Psi\equiv \tau + \H\,(2v -\sigma^\prime) \, , \qquad
\Phi \equiv \psi -2 v^\prime +
\sigma^{\prime \prime} - \H\, (2v -\sigma^\prime) \, ,\\
& W_i = u_i-s_i^\prime \, ,
\end{split}
\ee
while the transverse-traceless spin two field $\chi_{ij}$ is already
gauge invariant. It is also convenient to define the field
$\Sigma=\sigma/\Delta$.

\pagebreak[3]

We will couple the gravitational fields to a conserved\footnote{EMT
  conservation is not strictly required in massive gravity. A study of
  non-conserved EMT in FP can be found in \cite{Ford:1980up}.} energy-momentum
tensor $T_{\mu\nu}$,
\be
%\nonumber
S_T=-\int \di^4 x \, a^2\, h_{\mu\nu}\,\eta^{\m\a}\eta^{\n\beta}T_{\a\b} =-\int\di^4 x\  a^{2}
\left(\chi_{ij} \, T_{ij}+\Phi T_{00}-2 \, T_{0i} \, W_i+
\Psi \, T_{ii}\right),
\ee
with $\bar g^{\alpha \nu} \overline{\nabla}_{\alpha} T_{\mu\nu}=0 $,
where $\overline{\nabla}$ is the covariant derivative associated to
the background metric. The field $\Phi$ is the generalization of
Newtonian potential around the source in the linearized approximation.
For FRW, the EMT conservation is equivalent to
\be
T_{00}'=\de_i T_{0i} - \H \, \left( T_{00} + T_{ij} \delta_{ij} \right)  \,  , \qquad
\de_j T_{ij} = T_{i0}'+ 2 \H \, T_{i0} =a^{-2}\big(a^{2}T_{i0})'\, .
\ee

%%%%%%%%%%%%%%%%%%%%%%%%%%%%%%%%%%%%%%%%%%%%%%%%%%%%%%%
\section{Stable perturbations in curved backgrounds?}
%%%%%%%%%%%%%%%%%%%%%%%%%%%%%%%%%%%%%%%%%%%%%%%%%%%%%%%
\label{sec:phases}

Our goal is to study the dynamics of perturbations in curved backgrounds
and determine when one can get a theory free of instabilities. These
instabilities can be of {\em ghost} or {\em Jeans} type. The {\em
  ghost-like} instabilities are related to an infinite phase-space
volume. If they are present, the decay rate of the perturbative vacuum
will be infinite unless a cutoff is introduced in the theory
\cite{Cline:2003gs} (see also \cite{Dubovsky:2005xd}). In the Lorentz
breaking case, the different masses provide a natural energy scale to
place the cutoff. If we admit a hierarchy inside the mass scales we
may freeze the non-stable degrees of freedom, while still keeping some
of the masses below the cutoff. In an expanding universe, there is
also another important dimensional parameter, $H$. We will focus on
modifications such that at least some of the mass scales are inside
the horizon scale $m_i\gg H$. In this case, there is a natural
hierarchy inside the set of dimensional parameters, which allows to
define a large momentum cutoff $\L_c$ keeping the masses small,
\be
|\Delta| \leq m^2\left(\frac{m}{H}\right)^\a\sim \L_c^2,
\ee
with $\a > 0$.
Even if the addition of a cutoff may unveil phenomenologically acceptable
phases, to keep the discussion simple we will consider theories free from
{\em ghost-like} instabilities in the quadratic Lagrangian (see however~\cite{Blas:2008uz,Blas:2007zz}).

{\em Jeans-like} instabilities can be also present. By this we mean
instabilities that appear in a certain finite range of momenta. They
are the signature of the growth with time of the perturbation at
certain scales, and may even be interesting phenomenologically as a
contribution to the clustering of matter at large distances (see,
e.g.\ \cite{Libanov:2008mk}). Furthermore, in an expanding
universe, they may be settled beyond the horizon, where they are
presumably frozen. Again, we leave the study of this possibility for
future research \cite{BCNP}, and concentrate on lagrangians with a
stable spectrum.

\pagebreak[3]

It is also important to recall that in a FRW universe, energy is not a conserved
quantity, and its positivity does not guarantee stability. Nevertheless,
for scales smaller than the horizon, we can still use the positivity of the
energy associated to the conformal time as a {\em necessary} requirement
for stability (see also \cite{Abbott:1981ff,Deser:2001wx}).
As we will focus on these length scales, we will not discuss any global issue.
\smallskip

Finally, there are two concerns for massive gravity beyond the linear
theory~\cite{ArkaniHamed:2002sp}. The first one is the \emph{strong
  coupling} that emerges when one or more propagating states have their
kinetic terms suppressed by a small parameter.
%in a range of energies. 
In this case, the range of validity of the linear theory is
drastically reduced. Furthermore, if we consider the action
(\ref{ugac}) as an effective action with a cutoff, one expects the
contributions of higher order operators to become important at much
lower energies than the initial cutoff scale. To study this behaviour
one should analyze the scaling of relevant interaction
terms~\cite{Polchinski:1992ed} which is beyond the scope of this
paper. Here we just point out that when Lorentz invariance is
violated, the strong coupling cutoff can be present in energy and/or
momentum independently. We will accordingly speak of \emph{time} and
\emph{space} cutoff $\L_t$, $\L_s$, by making canonical the relative
quadratic terms in the action.

Besides, when the functions $m_i$ satisfy certain conditions, there is
a \emph{reduction of the phase-space}, i.e.\ not all the six DOF
of the gravitational perturbations propagate. It turns out that in
Minkowski those are the only ghost-free possibilities
\cite{Dubovsky:2004sg}. In general, unless there exists a symmetry
that enforces them (see for example the case of
bigravity~\cite{Berezhiani:2007zf}, or the case described
in~\cite{Dubovsky:2004ud}) these conditions are only satisfied for the
quadratic Lagrangian in very finely tuned backgrounds. This means that 
the analysis is very sensitive to small changes in the background
and probably to the interaction terms and higher order
operators~\cite{Rubakov:2008nh,Dubovsky:2004sg,Creminelli:2005qk}.
In the following we show how the generalization of LB massive gravity to curved
backgrounds is useful to circumvent both concerns.

\smallskip

In the next sections we analyze the spectrum of tensor, vector and
scalar perturbations, that at linearized level are not coupled, by
SO(3) symmetry.

%%%%%%%%%%%%%%%%%%%%%%%%%%%%%%%%%%%%%%%%%%%%%%%%%%%
\section{Tensor Modes}
%%%%%%%%%%%%%%%%%%%%%%%%%%%%%%%%%%%%%%%%%%%%%%%%%%%
\label{sec:tensors}

The action for the tensor perturbations is
\be
S^{(T)}=
\frac{M_{P}^2}{4}\int \di^4 x\ a^2\Big[
  -\eta^{\m\n}\pd_\m \chi_{ij}\pd_\n \chi_{ij}-a^2\,m_2^2\chi_{ij}\chi_{ij}
  \Big],\label{Stensor}
\ee
from which the EOM read
\be
\label{eq:tensorEoM}
 \eta^{\m\n}\pd_\m\pd_\n \chi_{ij} - 2\H \chi_{ij}^\prime - a^2  m_2^2 \chi_{ij} = 0\,.
\ee
The absence of tachyonic instabilities requires
\be
m_2^2\geq 0\,.
\ee
One can also readily check from (\ref{Stensor}) that there are no
ghost or gradient instabilities.\footnote{In an expanding universe, the
  friction term appearing in (\ref{eq:tensorEoM}) implies that the
 perturbation is frozen at large distance.  When imposing $m_2^2\geq
  0$, we are assuming that  the mass scale is well inside the
 horizon and unless otherwise stated, we will assume this to be 
 the case.}

\pagebreak[3]

\section{Vector Modes}
\label{sec:vectors}

Extracting the vector part from (\ref{eq:2nd}) we get
\be
S^{(v)}=\frac{M_P^2}{2}\int \di^4 x\ a^2\Big\{
  -(u_i- s_i')\Delta(u_i- s_i')+  a^2\big[ m_1^2\,u_i u_i+m_2^2\,s_j\Delta s_j
\big] \Big\}\,.
\label{eq:Svect}
\ee
The field $u_i$ is not dynamical and it can be integrated out
through its equation of motion,
\be
\Delta (u_i- s_i')-a^2 m_1^2\,u_i=0\,,
\ee
to yield
\be
S^{(v)}=\frac{M_P^2}{2}\int\di^4 x \ a^4\left[m_1^2  s_i' \frac{\Delta}{\Delta-a^2 m_1^2}
 s_i'+m_2^2 s_i\Delta s_i\right ].
\label{actV}
\ee
Therefore, the dispersion relation of the vector field $s_i$ breaks Lorentz invariance
at any scale (provided $m_1\neq m_2$). The action is free from
instabilities for
\be
m_1^2\geq 0\qquad \text{and}\qquad m_2^2\geq 0,
\ee
in complete analogy to what happens in Minkowski space. 

The case \underline{$m_1=0$} is particularly interesting, as it implies the
cancellation of the time-derivative term in (\ref{actV}), so that
there is no propagating vector mode. As we will see in
section~\ref{sec:m1}, this case is also important for the
scalar sector.

The canonically normalized field $s_i^c$ can be defined by a
rescaling:
\be
\label{eq:Vcanon}
s_i^c\equiv  \L_v(\eta)^{2}s_i\,,\qquad  \L_v(\eta)^{2}={a^2\,m_1 M_P}
\sqrt{\frac{\Delta}{\Delta-a^2 m_1^2}}\,,
\ee
with the action
\be
S^{(v)}=\frac{1}{2}\int\di^4 x \ \left\{ (s^c_i)' ( s^c_i)'
+\frac{m_2^2}{m_1^2}s_i^c\Delta s^c_i-
\Big[a^2 m_2^2+ \L_v^2(\L_v^{-2})''\Big] s_i^c s^c_i\right\}.
\label{eq:actVcan}
\ee
Therefore the canonical field has a LB dispersion relation and a time
dependent mass.

{}From (\ref{eq:Vcanon}) we can also read the naive \emph{temporal} strong
coupling scale of the vectors, that is momentum (and time) dependent. At
large momenta, $|\Delta|>m_1^2a^2$, we expect the {physical} strong
coupling scale to be given by $\L_t=\L_{v}/a\sim \sqrt{m_1
  M_P}$, that is also the expected cutoff for a gauge theory explicitly
broken by a mass term $m$. This implies that the theory can be
trusted only if the horizon scale does not exceed the cutoff,
$H<\sqrt{m_1M_P}$. In section~\ref{sec:conclusions} we will comment on
some physical consequences of this bound. A similar \emph{spatial}
strong coupling scale can be defined by making canonical the spatial
gradients.

%%%%%%%%%%%%%%%%%%%%%%%%%%%%%%%%%%%%%%%%%%%%%%%%%%%
\section{Scalar Modes}
\label{sec:Scalars}
%%%%%%%%%%%%%%%%%%%%%%%%%%%%%%%%%%%%%%%%%%%%%%%%%%%

The scalar sector of the theory is the most interesting one. In flat space
with the FP Lorentz invariant mass term it is a scalar mode which has the
lowest strong coupling scale, and it is this sector that shows crucial
differences when Lorentz symmetry is violated or spacetime is
curved~\cite{Rubakov:2004eb,vDVZAdS}. As we will see later, it is also here
that the difference between the maximally symmetric spacetimes and generic
FRW spaces arises. The analysis will show that generically there are two
scalar degrees of freedom, while the only possibilities with less degrees
of freedom are $m_1=0$ or $m_0=0$.

The scalar part of (\ref{eq:2nd}) can be written as (modulo total
derivatives)
\ba
\label{eq:actS}
S^{(s)}&=&\frac{M_P^2}{4}\int \!\di^4 x \,a^2\Big\{-6(\tau'+\H \psi)^2+
2(2\psi-\tau)\Delta \tau+4(\tau'+\H \psi)\Delta (2v- \s')\nonumber\\
&&\qquad\qquad\qquad+a^2\Big[m_0^2 \psi^2-2 m_1^2v\Delta v-m_2^2(\s \Delta^2\s+2\tau\Delta\s+3\tau^2)\\
&&\qquad\qquad\qquad\qquad+m_3^2(\Delta\s+3\tau)^2-2m_4^2\psi(\Delta\s+3\tau)\Big]\Big\}\,,\nonumber
\ea
where the $H$, $H'$ terms have canceled as promised.

In the de Sitter background (dS), when all  masses
are set to zero, the action reduces to the first line of (\ref{eq:actS}) and it is 
gauge invariant.\footnote{The first line of  (\ref{eq:actS}) differs from the
  standard action of the graviton in a FRW background by a term
 proportional to $(\H'-\H^2)\psi^2$ which cancels in dS
  (cf.\ \cite{Mukhanov:1990me}).} As previously remarked, for a FRW
background and vanishing masses the action is invariant only under
longitudinal spatial diffs (only $\sigma$ is undetermined).

From (\ref{eq:actS}) it is clear that $\psi$ and $v$ are Lagrange
multipliers enforcing the following constraints
\be
\label{eq:constr}
\begin{split}
&\psi = \frac{m_1^2 m_4^2 \left(\S+3 \tau \right) a^3 + 2 H m_1^2
\left(\S'+3 \tau'\right) a^2-2 \Delta m_1^2 \tau a-8 H
   \Delta \tau'}{8 H^2a\Delta +\left(m_0^2-6 H^2\right) m_1^2a^3}\,,  \\
& v= \frac{2 H a^2 m_4^2 \left(\S +3 \tau\right)
+4 H^2 a\S'+2 m_0^2 a\tau'-4 H \Delta\tau }{8 H^2a\Delta +\left(m_0^2-6 H^2\right) m_1^2a^3}  \, .
\end{split}
\ee
Notice that the behavior of these fields in FRW is qualitatively different
from Minkowski space. In particular, whereas in flat space, the cases
$m_0=0$ and $m_1=0$ are singular and must be treated separately, in
curved spacetime, $\psi$ and $v$ are always determined by equations
(\ref{eq:constr}).
After integrating out $v$ and $\psi$ we are left with a Lagrangian for
${\varphi} =(\S, \, \tau)^t$:
\be
\LL_{\S,\tau} = \frac12{\varphi'}^t \, \K \, {\varphi'}
           + \, {\varphi}^t \, \B \, {\varphi'}
           - \frac12{\varphi}^t \, \A \, {\varphi} \,,
\label{eq:acteff}
\ee
where
\begin{gather}
\K=
\frac{-M_P^2 a^2}{8 H^2\Delta +\left(m_0^2-6 H^2\right) m_1^2a^2}
\begin{pmatrix}
2H^2 a^2 m_1^2 & \quad   a^2m_0^2 m_1^2 \\[1ex]
 a^2m_0^2 m_1^2 &   \quad   m_0^2 \left(3 a^2 m_1^2- 4\Delta\right)
\end{pmatrix},\\[1em]
\B =
\frac{ M_P^2  \, \Delta H   \left(m_1^2-2 m_4^2\right)}
      {8 H^2\Delta +\left(m_0^2-6 H^2\right) m_1^2a^2}
 \, \begin{pmatrix} \,0  & \,1\, \\ -1 & \, 0\, \end{pmatrix}.
\end{gather}
We will first study the dynamics of (\ref{eq:acteff}) through the
Hamiltonian, for which the explicit expression of the matrix $\A$ is not
needed\footnote{As the form of this matrix in the general case is quite
  cumbersome and not particularly illuminating, we will not write it
  explicitly in this work.}. The conjugate momenta $\pi$ are
\be
\pi_i = \frac{\de \LL}{\de \varphi_i^{\prime}} =
 \K_{ij} \varphi_j^\prime -\B_{ij} \varphi_j  \, .
\ee
Thus, two DOF will propagate when the matrix $\K$ is non-degenerate,
i.e. when
\be 
\det ||\K|| \propto m_0\,m_1\neq0\,.
\ee

In this case one can express the velocities in terms of momenta, and
the resulting Hamiltonian is:
\be
\H_{\S,\tau} = \frac{1}{2}\pi^t \K^{-1} \pi
+  \frac12 \varphi^t \M \varphi\, ,\qquad \M=\left( \A +  \B \K^{-1}\B \right)\,,
\label{eq:ham}
\ee
with a rather simple kinetic term:
\be
\label{eq:kma}
{\mathcal K}^{-1} = \frac{1}{M_P^2a^2}
\begin{pmatrix}
\displaystyle3-\frac{4\Delta}{a^2m_1^2}\ \  & -2 \\
 -2\  & \displaystyle\frac{2H^2}{m_0^2}
\end{pmatrix} .
\ee
The theory is free of ghosts when the kinetic energy matrix $\K^{-1}$
is positive definite, that translates into the following conditions:
\be\label{nog}
m_1^2>0\,, \qquad  0<m_0^2\leq6H^2\,.
\ee
Therefore contrary to the flat space case, we can still have a well
defined kinetic term with two propagating degrees of freedom. In fact
a window for $m_0^2$ opens up, and this allows even for a ``non-FP''
Lorentz-invariant mass term free of ghosts (and vDVZ discontinuity,
see section~\ref{sec:vDVZ}).\footnote{Recently, non-Fierz-Pauli
  lagrangians with scale dependent masses were also considered in
  \cite{Dvali:2008em}. Notice, though, that in that case Lorentz
  invariance made the masses depend on both space and time, whereas in
  this work we are dealing only with time dependent masses.}

It is also instructive to look at the no-ghost conditions in the low and
high momentum regimes. We find,
\be {\rm no \ ghost}
\begin{cases}
\text{at large momenta:}&m_1^2>0\,, \quad m_0^2>0\\
\text{at small momenta:}&m_1^2>0\,,\quad 0<m_0^2\leq6H^2\,.
\end{cases}
\ee
Therefore a nonzero curvature allows the scenario where the theory is free
of ghosts in the ultraviolet but there is one ghost mode at large
wavelengths; this happens for $m_{1}^2>0$ and $m_0^2>6H^2$. Such a ghost
mode at very large distances would not necessarily render the theory
phenomenologically sick, but would indicate a large scale instability of
backgrounds with curvature smaller than $m_0^2/6$, including the limiting
case of Minkowski ({\em Jeans-like} instability, in the language of the
section~\ref{sec:phases}).

From (\ref{eq:kma}) one can find when the scalar sector suffers from
strong coupling due to a small kinetic term. As happens for the vector
modes, one of the strong coupling scales is related to the smallness
of $m_1$, whereas the other one depends on the ratio $m_0^2/H^2$. When
this ratio is not small (and compatible with the ghost-free condition
(\ref{nog})), both the scalar and vector sector become strongly
coupled at the same \emph{time} scale $\Lambda_t\sim \sqrt{m_1 M_P}$.

\medskip

The analysis of the positivity of the ``mass'' term $\M$ is rather
cumbersome and we will consider just the high momentum limit (larger
than the rest of the scales: $m_i$, $H$ and $H'$). In this case,
requiring that the mass matrix in (\ref{eq:ham}) is positive definite
gives
\be
\label{eq:generalgrad}
m_3^2-m_2^2< \frac{\left(m_1^2-2 m_4^2 \right)^2}{16m_0^2}\,,
\qquad
H'a^{-1}<-\left[\frac{m_1^2}{4}+\frac{\left(m_1^2-2 m_4^2 \right)^2}{16m_1^2}\right],
\ee
(where we have used $m_1^2>0, \, m_0^2>0$ and $H^2>0$.) When the
previous conditions are satisfied there is no gradient instability at
small distances. Notice that the r.h.s.\ in the last condition is
always negative, meaning that only a FRW background with an expanding
horizon can be stable. Besides, one can easily check that the previous
conditions are inconsistent in the Lorentz-invariant case.

\medskip

To summarize, in the \emph{non degenerate case} of $m_{0,1}\neq0$, we found
two DOF where
\begin{itemize}
\item there is no ghost provided $m_1^2>0$, $6H^2\geq m_0^2>0$ 
\item
  there is no gradient instability when (\ref{eq:generalgrad})
  are satisfied (so $H'$ is negative). 
\end{itemize}
The difference that we found with the maximally symmetric case, where
there is necessarily a gradient instability, implies the presence of a
\emph{spatial} strong coupling problem in this limit. In fact in
approaching the dS background the spectrum of the Hamiltonian must
pass through the case in which one of the modes is frozen, because the
determinant of ${\cal M}$ vanishes and accordingly the ``spatial''
part of its dispersion relation will
vanish.\footnote{See~\cite{Dubovsky:2004sg} for a discussion of the
  modifications to these dispersion relations coming from higher order
  operators.}

\medskip

In the degenerate cases, when $m_0$ or $m_1$ vanish, there are less
DOF and a separate analysis is given in the following sections.  The
$m_0=0$ case is related to the Fierz-Pauli case~\cite{Fierz:1939ix},
whereas the case $m_1=0$ appears naturally in the ghost condensate and
bigravity theories \cite{Dubovsky:2004ud,Blas:2007zz,Berezhiani:2007zf}.

%%%%%%%%%%%%%%%%%%%%%%%%%%%%%%%%%%%%%%%%%%%%%%%%%%%%%%%%
\subsection{\bf The phase $m_0=0$}\label{sec:m0g}
%%%%%%%%%%%%%%%%%%%%%%%%%%%%%%%%%%%%%%%%%%%%%%%%%%%%%%%%

For $m_0=0$, the field $\tau$ is an auxiliary field as one can check from
the action (\ref{eq:acteff}). Even if there is only one remaining DOF, the
general treatment is quite involved and it is presented in
appendix~\ref{app1}. In this section we will just state the results and
study some particular cases.

The EOM for $\t$ yield the constraint (\ref{eq:tau}), which once once
substituted in the action gives a (quite complicated) effective Lagrangian
for $\Sigma$. Its kinetic part is
\be\label{kinm0}
 \frac{\cal K}{ a^4M_P^2} = \frac{3a^3m_1^2
\left[a (m_4^2)^2 + 2 (a m_\mu^2H^2-H (m_4^2)'+m_4^2 H') \right]
-4[am_4^2 (m_1^2-m_4^2)+m_1^2H']\Delta}
{a m_1^2(2\Delta-3a^2m_4^2)^2-2(4\Delta-3a^2 m_1^2)[3 a^2H( aH m_\mu^2- (m_4^2)')
-\left(2 \Delta
   -3 a^2 m^2_4\right) H']} ,
\ee
where $m_\mu^2=3(m_3^2-m_4^2)-m_2^2$. The positivity of the kinetic energy
(no ghost) for large momenta gives
\be
\label{noghostm0}
\frac{a  m_4^2 (m_1^2- m_4^2)+m_1^2  H'}{ a m_1^2+4 H'} >0 \, .
\ee
On the other hand at small momenta the kinetic term reduces to
$${\mathcal K}|_{\Delta=0}= a^2 M_P^2/3,$$
which is, remarkably, always positive.

One can show (cf.\ appendix~\ref{app1}) that ${\mathcal K}$ is positive also
at any intermediate momenta in the variable $\Delta$ provided that, in
addition to (\ref{noghostm0}), one has
\be
\label{rootnum}
m_1^2\geq 0\,,\qquad \left(\frac{a(m_4^4+2 H^2 m_\m^2)+2 m_4^2 H'-2 H (m_4^2)'}{a m_4^2(m_1^2-m_4^2)+m_1^2 H'}\right)>0\,.
\ee
When these conditions are saturated we are led to a case 
with a vanishing kinetic term, as discussed below (see (\ref{eq:partmass})). 
Due to its analogy with the special case discussed in \cite{Deser:2001wx},
we will refer to this case as \emph{partially massless}.

The condition (\ref{rootnum})
refers to modes at large distances (eventually outside the horizon) and is
not present in the Minkowski spacetime. In dS, taking the Lorentz-invariant
FP limit (\ref{eq:PFlimit}) with $m^ 2\equiv\b=-\a$, the previous conditions
reduce to the Higuchi bound \cite{Higuchi:1986py},
\be
2H^2\leq m^2.
\ee  
Contrary to this case, the LB mass terms allow for a unitary massless
limit. In fact, if the mass is an appropriate function of the
conformal time, this limit can be free from ghosts also in the
Lorentz-invariant case (see section~\ref{PFm0}).
 
{}From the same kinetic term we can also estimate the strong coupling scale
of the field $\Sigma$, since the canonical field $\S^c$ is defined at high
momentum by the rescaling
\be
\Sigma^c=\L_\Sigma\S \,,\qquad \L_\S = a\frac{M_Pm_1}{\sqrt{-\Delta}}
          \left[\frac{am_1^2\, x (1- x)+ H'}{am_1^2+4 H'}\right]^{\frac12},
\ee
where $x=m_4^2/m_1^2$.

\medskip

Concerning the potential term $\M$, it can be written as
\be
\label{eq:m0pot}
\M=\frac
  {m_2^2\,b^2+c\Delta+d\Delta^2+e\Delta^3+(m_2^2-m_3^2)\Delta^4}
  {q^2}
\ee
where $b$, $c$, $d$, $e$ are functions of $m_i$ and $\H$, whereas $q$ is a
second order polynomial in $\Delta$.

The absence of gradient instabilities, equivalent to the positivity
of ${\cal M}$, requires in the ultraviolet and infrared regimes the
following simple conditions:
\be
\begin{cases}
\text{at large momenta} & m_2^2>m_3^2 \\
\text{at small momenta} & m_2^2 > 0 \,.
\end{cases}
\ee
We see that for $m_2^2\geq m_3^2$ the potential is free from gradient
instabilities that would be as dangerous as ghost instabilities as they
would imply an infinitely fast instability \cite{Dubovsky:2005xd}. Notice
also that at zero momentum, the condition $m_2^2>0$ required for the
stability of tensors and vectors, enforces positivity of the potential.
 This implies, together with the stability of the kinetic
term, that at small momentum the theory is always stable.

At intermediate scales, the analysis becomes very technical, and a method
to check for the positive definiteness is presented in the appendix
\ref{app1}.

%%%%%%%%%%%%%%%%%%%%%%%%%%%%%%%%%%%%%%%%%%%%%%%%%
\subsection{Particular cases with $m_0=0$}
%%%%%%%%%%%%%%%%%%%%%%%%%%%%%%%%%%%%%%%%%%%%%%%%%

Some interesting subcases of the $m_0=0$ dynamics can be found
looking at the numerator and denominator of eq (\ref{kinm0}). When 
the denominator vanishes, the field $\tau$ disappear from the EOM and
the constraint (\ref{eq:tau}) does not hold anymore.
The analysis of this situation is presented in section \ref{diffm0}.

When the mass parameters are fine tuned in such a way that the
numerator of (\ref{kinm0}) vanishes, the $\S$ field does not propagate
and it becomes an auxiliary field. This possibility is examined in
section~\ref{zerom0}, where we also show its relation to a gauge
invariance related to conformal invariance. Finally, the Fierz-Pauli
Lagrangian in dS is a subcase of the phase $m_0=0$ where this fine
tuning can occur. We study this possibility in section~\ref{PFm0}.

%%%%%%%%%%%%%%%%%%%%%%%%%%%%%%%%%%%%%%%%%%%%%%%%%
\subsubsection{Time diffeomorphisms in FRW}
\label{diffm0}
%%%%%%%%%%%%%%%%%%%%%%%%%%%%%%%%%%%%%%%%%%%%%%%%%

From the constraint (\ref{eq:tau}) for $\tau$ as a function of $\S$, we see
that it is singular for specific values of the masses. This happens
when
\be
\label{eq:timed}
 m_1^2=2m_4^2=-4a^{-1}H'\,,\qquad
 m_\m^2=\frac{ a (H^2)'-2 H''}{a^2 H}.
\ee
(If only the first condition holds, the constraint for $\tau$
 (\ref{eq:tau}) reduces to $\t=-\S/3$.)
 
\pagebreak[3]

In this case, and away from dS (we are assuming $m_1\neq 0$), the
field $\tau$ does not appear at all in the action. This corresponds to a
restoration of the gauge symmetry corresponding to time diffeomorphisms,
$\delta(2v-\s')=2\xi_0$.

The final Lagrangian in terms of $\S$ is then given by
\be
\label{eq:sactionm0I}
\mathcal L=\frac{M_p^2a^2}{2}\left\{\frac{a
    H'}{\Delta+3aH'}\S'^2-\frac13\left[m_2^2+
\Delta\frac{3 (aH')^2 + (2aH' +H''/H)\Delta}{(\Delta + 3 aH')^2}\right]\S^2\right\}.
\ee
Notice that in Minkowski this phase has $m_0=m_1=m_4=0$ and features
an enhanced gauge invariance mentioned
in~\cite{Berezhiani:2007zf}.  In contrast to the case of Minkowski,
and to the case of dS, in FRW background the field $\S$
propagates.  It is also clear from (\ref{eq:sactionm0I}) that the
kinetic term is positive definite provided that $H'<0$. 
Concerning the potential term, let us consider scales well inside the
horizon.  The requirement of positive energy at these scales gives the condition
\be
m_2^2\geq -(2aH' +H''/H)\,,
\label{eq:partcond}
\ee

When inequality (\ref{eq:partcond}) is exactly saturated, 
the scalar degree of freedom has vanishing speed in the
at high momenta.  Its dispersion relation is then $\omega^2\simeq const. +
1/\Delta$.

Finally, let us note that if for $H'<0$ the scalar mode is well
behaved, the limit of vanishing $H'$ leads to a vanishing kinetic term
and thus to strong coupling once interactions are taken
into account. The resulting \emph{time} strong coupling scale can
be estimated as
\be
\L_s=a\, M_P \sqrt\frac{a H' }{\Delta +3 aH'}\,.
\ee
that is clearly more dangerous at short distances $|\Delta|\gg|aH'|$,
where it may become sensibly lower than $M_P$.

%%%%%%%%%%%%%%%%%%%%%%%%%%%%%%%%%%%%%%%%%%%%%%%%%%%%%
 \subsubsection{Partially massless}\label{zerom0}
 %%%%%%%%%%%%%%%%%%%%%%%%%%%%%%%%%%%%%%%%%%%%%%%%%%%%%

Another particular case appears when, after integrating out $\tau$, the
kinetic term of~$\S$ cancels. This happens when the inequalities
(\ref{noghostm0}) and (\ref{rootnum}) are saturated, 
\be
\label{eq:partmass}
m_1^2=\frac{a\, m_4^4}{a\,m_4^2+H'}\,,\qquad
a(m_4^4+2 H^2 m_\m^2)+2 m_4^2 H'-2 H (m_4^2)'=0.
%
%}-2m_4^2\H'+4\H m_4 m_4'}{6\H^2}.
%\nonumber
\ee
In the Fierz-Pauli limit this expression reduces to the partially
massless case of de Sitter space(cf.\ \cite{Deser:2001wx}) and
corresponds to a situation without propagating scalar degrees of
freedom. In the Lorentz invariant case with constant masses in de
Sitter space, this fact
 is related to a conformal invariance \cite{Deser:1983mm}. 
In the most general case, one can prove that the system is invariant under the transformation
\be
\delta \psi =-2(\xi'+\H \xi)+\phi_t,\quad  \delta v=-\xi+\zeta',\quad  \delta \t=2 \H\xi+\phi_s, \quad  \delta \s=2 \zeta,
\ee
with
\be
\begin{split}
&\xi=-\frac{a_4^2(a m_4^2\zeta+H'\zeta +H\zeta')}{H(a m_4^2+2H')},\quad \quad \phi_s=m_4^2 \zeta\\
&\quad \phi_t=\frac{\zeta m_2 (a m_4( m_4^2-4H^2)+2 m_4 H'-4 H m_4')}{2 a H^2},
\end{split}
\ee 
only when the extra condition 
\be
\label{eq:gaugepot}
a[m_4^4+2H^2(2m_2^2-3m_4^2)]+2 [m_4^2 H'-H (m_4^2)']=0
\ee
is satisfied. The previous condition implies the cancellation of the
potential part once (\ref{eq:partmass}) is satisfied. In the
Lorentz-invariant limit with constant masses and dS background
(\ref{eq:gaugepot}) is always satisfied when (\ref{eq:partmass})
holds. Notice also that the existence of this sort of scale invariance
is general even if the kinetic term is not invariant under diff away
from de Sitter.

%%%%%%%%%%%%%%%%%%%%%%%%%%%%%%%%%%%%%%%%%%%%%%%%%
\subsubsection{Lorentz-invariant FP limit with time dependent
  masses}\label{PFm0}
%%%%%%%%%%%%%%%%%%%%%%%%%%%%%%%%%%%%%%%%%%%%%%%%%

In the Fierz-Pauli limit ((\ref{eq:PFlimit}) with $m^
2=\beta=-\alpha$) the mode $\Sigma$ propagates. However, the
conditions (\ref{eq:partmass}) can be still be satisfied in dS (and
only for this background) provided that $m$ satisfies the differential
equation
\be
\label{eq:PFpartmass}
4 \H m'=(a^2m^2-2 \H^2)m\,.
\ee
This equation can be integrated to yield
\be
\label{eq:PFsol}
m^2(\eta)=\frac{2H^2 m_I^2}{m_I^2+(2H^2-m_I^2)a(\eta)}\,,
\ee
where $m_I$ is the value of the mass at the time corresponding to
$a(\eta)=1$. The resulting mass runs from $m_I^2$ to $2H^2$ when $a$
runs from 0 to 1. Notice that choosing the initial conditions
corresponding to a constant mass, $m_I^2=2H^2$, we recover the
partially massless case discussed in \cite{Deser:2001wx}. A similar
situation could be studied for the non-Fierz-Pauli (Lorentz-invariant)
case ($m_1=m_2\neq m_3=m_4$, $m_0=m_3-m_2$).

%%%%%%%%%%%%%%%%%%%%%%%%%%%%%%%%%%%%%%%%%%%%%%%%%
\subsection{\bf The phase  $m_1=0$}
\label{sec:m1}
%%%%%%%%%%%%%%%%%%%%%%%%%%%%%%%%%%%%%%%%%%%%%%

The case $m_1=0$ is particularly interesting in the Minkowski
background, as only the tensor modes propagate. As we will show, there is
a corresponding effect in  dS, while one scalar mode starts to propagate in a
FRW background.
 When $m_1=0$ the fields $\S$ is not dynamical as one can check in
action~(\ref{eq:acteff}). Accordingly, its EOM is
\be
\label{constIII}
\H(m_2^2-m_3^2)\S=m_4^2\tau'-\H(m_2^2-3m_3^2)\tau\,.
\ee
Notice that again the Minkowski space limit $\H=0$ is peculiar and the
degree of freedom associated to $\tau$ is not present\footnote{Also,
  the case $m_2=m_3$ should be treated differently.}.
In curved space, generically $\S$ is determined by (\ref{constIII})
and when it is substituted back in the action, after integration by
parts, yields the Lagrangian
\ba
\label{Lagrm1}
{\mathcal L}&=&\frac{M_P^2a^2}{H^2}
\left\{\frac{m_\eta^4}{2(m_2^2-m_3^2)}\tau'^2
-\left[\frac{H'}{a}\Delta
    +\frac{m_2^2[\H^2\left(m^2_2-3m^2_3+3 m^2_4\right)- m^2_4 H'a]}
          {m^2_2-m^2_3}\right.\right.\\
&&{}\qquad\qquad\qquad\qquad\qquad
          \left.\left.-\frac{\H\left[m^2_4 \left(m^2_3 (m^2_2)'-m^2_2 (m^2_3)'\right)+m^2_2 \left(m^2_3-m^2_2\right) (m^2_4)'\right]}{\left(m^2_2-m^2_3\right)^2}
\right]\tau^2\right\}.\nb
\ea
where $m_\eta^4=m_0^2(m_2^2-m_3^2)+m_4^4$. {}From the previous expression
we discover that in dS, the phase $m_1=0$ has no propagating degrees of
freedom (in the sense that the action is $\Delta$ independent so that there
is no dynamics in $\vec x$ space), even if, in comparison to the Minkowski
case, the scalar sector has a kinetic term from which we expect a ghost
condensate like dispersion relation coming from higher derivatives
\cite{Dubovsky:2004sg}. Besides, the potential strong coupling scales
 $\Lambda_s$ and $\Lambda_t$ are easily read out from the previous expression.

Thus, in general the phase $m_1=0$ is quite rich, and particularly simple.
Ghostlike instabilities are avoided imposing $m_\eta^4(m_2^2-m_3^2)\geq 0$.
To get rid of gradient instabilities in this case, it is enough to impose
$H'<0$, whereas the tachyon free condition can also be read from
(\ref{Lagrm1}). For the case with constant masses, it reduces to
$m_2^2[\H^2\left(m^2_2-3m^2_3+3 m^2_4\right)- m^2_4 H'a]\geq 0$.

%%%%%%%%%%%%%%%%%%%%%%%%%%%%%%%%%%%%%%%%%%%%%%%%%%%%%
\subsection{Particular cases with $m_1=0$}
%%%%%%%%%%%%%%%%%%%%%%%%%%%%%%%%%%%%%%%%%%%%%%%%%%%%%

A direct inspection of (\ref{Lagrm1}) and (\ref{constIII}) shows 
some interesting subcases for the mass parameters. First, when
the r.h.s.\ of eq. (\ref{constIII}) cancels, this equation is no longer a constraint
for $\S$.
Besides, for $m_\eta=0$, the kinetic term for $\t$ cancels in the action.
We devote the rest of this section to the analysis of these possibilities.

%%%%%%%%%%%%%%%%%%%%%%%%%%%%%%%%%%%%%%%%%%%%%%%%%%%%%
\subsubsection{The case $\mathbf{m_2^2=m_3^2}$} \label{m1den}
%%%%%%%%%%%%%%%%%%%%%%%%%%%%%%%%%%%%%%%%%%%%%%%%%%%%%
When $m_2^2 = m_3^2$ the kinetic term of $\tau$ is zero. In this case
$\tau$ is non-dynamical and can be eliminated from the action. The
only degree of freedom now is $\Sigma$ with a Lagrangian
\be
{\cal L}= \frac{a^4 M_P^2}{2}\frac{\left(6 m_2^2 m_4^4-9 m_4^6+4 m_0^2 m_2^4\right)m_4^4
}{\left(2 m_2^2-3 m_4^2\right)
\left(2 m_0^2 m_2^2-3 m_4^4\right)^2}
\left[\frac{  m_4^4 }{2 \left(2 m_2^2-3 m_4^2\right)   {\cal H}^2} \,  {\Sigma'}^2
%\right.\\&&\hspace{3cm}\left.
+m_2^2  \, \Sigma ^2
\right].
\ee
Again, this mode has no dynamics in space.
From direct inspection we can derive the  strong coupling scale,
and the region of parameters where this mode disappears.

%%%%%%%%%%%%%%%%%%%%%%%%%%%%%%%%%%%%%%%%%%%%%%%%%%%%%
\subsubsection{The case  $\mathbf{m_\eta =0}$}\label{m1num} 
%%%%%%%%%%%%%%%%%%%%%%%%%%%%%%%%%%%%%%%%%%%%%%%%%%%%%

Finally, for $m_\eta=0$ we are back to a situation without scalar
propagating degrees of freedom but  still with a potential part at the
linear level. In Minkowski also this part  vanishes and the  field $\tau$ is not
determined (indeed, there is an additional gauge invariance).
In dS, this happens when
\be
m_\eta=0\,,\qquad
\frac{\H(m^2_2-3m^2_3+3 m^2_4)-(m^2_4)'}{m^2_4}=
 \frac{m^2_3 (m^2_2)'-m^2_2 (m^2_3)'}{ m_2^2(m^2_2-m^2_3)},
\ee
and outside  this region of the parameter space,  the EOM gives $\tau= 0$.

%%%%%%%%%%%%%%%%%%%%%%%%%%%%%%%%%%%%%%%%
\section{Coupling to matter and vDVZ discontinuity}
\label{sec:vDVZ}
%%%%%%%%%%%%%%%%%%%%%%%%%%%%%%%%%%%%%%%%

Though the vDVZ discontinuity is one of the main phenomenological
difficulties of FP massive gravity in flat space, it is known that it may
be circumvented in curved backgrounds \cite{vDVZAdS} or when one considers
Lorentz violating mass terms \cite{Rubakov:2004eb}. For AdS or dS, the vDVZ
discontinuity is avoided by hiding the effects of the mass at distances
larger than the horizon, and as a consequence there is no modification of
gravity at scales smaller than the Hubble radius. In this section we will
see that some of the massive gravity phases we have studied  allow for a
modification of gravity at scales shorter than the horizon scale and still
compatible with GR at linear order. We will focus on the  gravitational
potentials produced by a ``point-like'' conserved source.

The tensor part is described by a massive graviton with mass given by
$m_2^2$. Phenomenologically, this mass is constraint by cosmological
and astrophysical observations (see e.g.\
\cite{Arun:2009pq,Dubovsky:2004ud}), and has no impact on the
gravitational potentials for point-like sources. Also vectors modes do
not affect these potentials (for cosmological constraints see
\cite{Bessada:2009qw}). For our purposes only scalar perturbations are
relevant.

Let us briefly review the situation of standard GR in presence of ``point
like'' conserved sources, in Minkowski or dS background:
\be
T_{00} =\frac{\rho(r)}{a} \, , \qquad T_{0i}=T_{ij}=0 \, .
\ee
In GR, there is no scalar propagating DOF and the gauge invariant
potentials are determined from the sources as
\ba
\Phi_{GR} &=&\Psi_{GR}=\frac{1}{M^2_P \,  \Delta} T_{00} \, .
\ea
Recall that the perturbations are defined with respect to a non-flat metric.
Thus, both the background and the perturbations play a role in the
gravitational dynamics around local sources.

As described in  section~\ref{sec:Scalars}, the generic massive
gravity case has two propagating DOF in the scalar sector. In this section
we are interested in static solutions in the presence of static sources.
More concretely, we will consider time scales short enough such that we can
consider the background metric constant\footnote{In this limit the standard
  Fourier analysis is well suited to analyze the EOM and energy is a
  conserved quantity. It is also clear that if the limit is not singular
  the results are equivalent to those of Minkowski space considered in
  \cite{Dubovsky:2004sg,Rubakov:2004eb}.}. By inspecting the EOM's in this
limit, time derivatives can be neglected provided that $\omega, \, \omega
H\ll k^2$; $\omega H \ll m^2_i$ and $\omega H \ll E^2$ where $\omega^{-1}$
is the typical time scale for the variation of the gravitational
perturbations and $E$ is the energy scale of the sources.

Once that time derivatives of the two dynamical fields $\S$ and $\tau$ are
neglected, and in the regime $H^2,H', m_i'\ll m_i^2\ll \Delta$, the EOM can
be solved in a straightforward though lengthy way. The generalization of
the Newtonian potential is the quantity $\Phi$ and we get
\be
\label{phi}
\Phi=\frac{  n_2 \Delta^2 +n_1 \Delta + n_0}
{d_3 \Delta^3+d_2 \Delta^2+d_1\Delta+ d_0} \, ,
\ee
where the $n_i$ and $d_i$ are polynomials in the masses. The physics
relevant for the vDVZ discontinuity is captured by expanding $\Phi$ in
powers of $1/\Delta$, e.g.\ $\Delta \gg m_i^2$.
\be
\begin{split}
\Phi & =  \frac{T_{ii}+T_{00}}{M_P^2 \Delta} - \frac{u \, T_{00}+ v \, 
T_{ii}}{2 M_p^2 \Delta^2 
\left(m_2^2-m_3^2\right)} + \, O\left(\frac{1}{\Delta^3} \right) \, , \\[.3cm]
& u = a^2 \left[ m_\eta^4 
+m_2^2\left(6 m_3^2-4 m_4^2-2m_2^2\right) \right] \, ,\\ 
&v =  a^2 \left[ m_\eta^4 -2 m_2^2m_4^2 \right] \, . \\
\end{split}
\label{eq:phi}
\ee 
Thus, at small distances we get the GR result plus
corrections.\footnote{The expression (\ref{eq:phi}) is valid for
  distances smaller than the inverse of mass. For distances of the
  order of the inverse of the mass, the appearance of a pole in
  (\ref{phi}) makes the series ill defined. The exact solution can be
  easily found and one can see that the perturbations acquire a Yukawa
  tail. Thus, this modification of Newtonian potential has the
  desirable feature of keeping the perturbations small at large
  distances.} Also $\Psi$, that is important for post-Newtonian tests,
has the same structure:
\be
\Psi  =  \frac{T_{00}}{M_P^2 \, \Delta } -
\frac{a^2}{2M^2_P\,\Delta^2}\left[T_{00} \frac{ m_\eta^4 -2 m_2^2
m_4^2}{m_2^2-m_3^2}+ T_{ii}\frac{ m_\eta^4} { m_2^2-m_3^2} \right]
+O\left(\frac{1}{\Delta^3}\right) \, .
\label{eq:psi}
\ee
Clearly, no discontinuity is present at small distances provided that
$m_2^2\neq m_3^2$ (notice also that $m_1$ has disappeared from the
previous expression). When $m_2^2 = m_3^2$, the previous expressions
are not valid and a discontinuity is present, as it can be 
established by noting that in the UV the EOM imply 
\be
2m_3^2\Psi=m_4^2\Phi\,,
\ee
which does not hold in GR.

%%%%%%%%%%%%%%%%%%%%%%%%%%%%%%%%%%%%%%%%%%%%%%%%%
\subsection{Coupling to matter for $m_1 =0$}\label{m1vd}
%%%%%%%%%%%%%%%%%%%%%%%%%%%%%%%%%%%%%%%%%%%%%%%%%

The case $m_1=0$ is of particular interest, as in flat space there is
no scalar DOF and the potential features a correction linear with $r$,
invalidating the linearized approximation at large distances. In a
curved space the scalar $\tau$ propagates and the gauge invariant
potentials $\Phi$ and $\Psi$ can be written as a combination of the
source, $\tau$ and its time derivatives as
\be
\begin{split}
&\Psi = \Psi_{GR} +a\left( \frac{2a\, H m_2^2 m_4^2 \,\tau + m_\eta^4 \,\t'}{2\Delta H(m_2^2-m_3^2)}
\right),\\
&\Phi=\Psi+a \,m_2^2\left(\frac{2a\, H (m_2^2-3m_3^2) \,\tau -m_4^4 \,\t'}{\Delta H(m_2^2-m_3^2)}\right).
\end{split}
\label{eq:pot1}
\ee
Here we have used the expression for $\t''$ obtained from the EOM, namely:
\be
\label{eq:taup}
\t''=\frac{2(m_2^2-m_3^2) H'}{a\, m_\eta^4 M_P^2}(T_{00}-M_P^2\Delta \tau)+q_1(m_i,H)
 \,\tau+ q_2(m_i,H) \,\tau'\,,
\ee 
where $q_{1,2}$ are functions of the background and the masses, finite
in the limit $m_i\rightarrow 0$.  From these expressions one can study
the behavior of potentials in the limit $m_i\to0$.

First, consider the dS background, $H'=0$. In this case, the first
term in the r.h.s.\ of~(\ref{eq:taup}) vanishes. As a result, the only
particular solution (vanishing for zero sources) is $\tau =0$, and the
potentials (\ref{eq:pot1}) coincide with those of GR.  Remarkably,
also the linear term appearing in Minkowski~\cite{Dubovsky:2004sg} is
absent in dS.

A similar situation happens in FRW background: when $H'\neq0$, the
first term in the r.h.s.\ dominates in the $m_i\to0$ limit, and $\tau$
remains finite:
\be
\tau\sim \frac{T_{00}}{\Delta  M_P^2}+O(m^2)\,.
\ee
This implies that the corrections to $\Phi$ and $\Psi$ with respect to 
GR vanish in this limit, and there is no vDVZ discontinuity.

Further insight can be gained by looking at explicit solutions of
(\ref{eq:taup}). These can be found by assuming a special time
dependence of the masses and the scale factor:
\be
a = \left(\frac{\eta}{\eta_0} \right)^\ell \, , \qquad m_i^2(\eta) = a^{s} \lambda_i \, ,
\label{eq:ans}
\ee
where $\lambda_i$ are constants of dimension two. The EOM for $\tau$
(\ref{eq:taup}) then reduces to
\ba
%\begin{split}
&\displaystyle\tau''+ \frac{2 + \ell(4+s) }{\eta} \tau' +
\frac{2 a^{-s-3}  \ell (\ell+1)  \rho(r) (\lambda_2 - \lambda_3)}{  M_P^2 \lambda_\eta^2 
 \eta^2} +
 \frac{2a^{-s-2} \, \ell}{\lambda_\eta^2}
\left ( -(\ell +1) (\lambda_2 - \lambda_3) \Delta 
+\right.
\nonumber\\
&\displaystyle\left.
 a^{2+s} \lambda_2 \left[
\lambda_4 + \ell (\lambda_2 -3 \lambda_3) + \lambda_4 \ell (s+4) \right] \right ) \tau
=0\,,
%\end{split}
\label{eq:tau1}
\ea
where $\lambda_\eta^2 = \lambda_4^2 + \lambda_0 ( \lambda_2 - \lambda_3)$.

\pagebreak[3]

For a general background one can easily find an exact solution of
(\ref{eq:tau1}) when $s=-2$. The solution that is relevant to us can be
written as\footnote{\label{foot}The general solution is of the form
$$\tau=t^ {-1/2-\ell}\left( C_1 t^{\b[\Delta]/2}+C_2t^{-\b[\Delta]/2}\right)+\tau_P.$$ 
Stability requires $|\b(\Delta)|-(2l+1)<0$, which at high energies implies
$l(l+1)\l_\eta^ 2(\l_2-\l_3)\propto -H'\l_\eta^ 2(\l_2-\l_3)>0$. This
condition was readily derived in section~\ref{sec:m1} from direct
inspection of the Lagrangian. One can also check that the solution is
stable at any scale if in addition $l>-1/2$ and
$\l_2\l_\eta^2\{\l_4+l(\l_2-3\l_2+2\l_4)\}<0$.}
\be
\label{eq:mu}
\tau_P = \frac{T_{00}}{M_P^2(\Delta-\m^2)},\quad \quad \ \m^2=\frac{
2 \l_2\l_4(1+2\ell)-(\ell+1)\l_\eta^ 2
 + 2\ell (\l_2^ 2-3\l_2\l_3)}{2 (\ell+1)(\l_2-\l_3)}.
\ee
In this case, the potentials (\ref{eq:pot1}) are
\be
\begin{split}
\Psi = \Psi_{GR}+ \left(\frac{2\l_2\l_4-\l_\eta^2}{2\Delta(\l_2-\l_3)}\right)\t_P,
\quad \Phi=\Psi+    \l_2\left(\frac{\l_2-3\l_3+\l_4}{\Delta(\l_2-\l_3)}\right)\t_P.
   \end{split}
\ee
and one can check that there is no discontinuity in the massless
limit. As recalled in appendix~\ref{app2} one can work out the
explicit expression for the potentials in position space to get (we
assume $\m^2>0$)
\be
\label{eq:potm1}
\begin{split}
&\Psi= \Psi_{GR}\left[
1+ \left(\frac{2\l_2\l_4-\l_\eta^2}{2\m^2(\l_2-\l_3)}\right)\left(e^{-\m r}-1\right)
\right], \\ 
&\Phi= \Phi_{GR}\left[
1+ \left(\frac{2\l_2^2-6\l_2\l_3+4\l_2\l_4-\l_\eta^2}{2\m^2(\l_2-\l_3)}
\right)\left(e^{-\m r}-1\right)
\right],
\end{split}
\ee
where $\m$ can be read from (\ref{eq:mu}). This result differs from
the one found in flat space (see e.g.\ \cite{Dubovsky:2005dw}) in
some essential facts: first, instead of the linear correction to
$\Phi$ that appears in Minkowski, we found an exponential function
that decays to a constant at large values of $r$ ($r\gg\m^{-1}$). This
is an infrared modification of GR whose magnitude depends on a ratio
of masses, i.e.\ it gives finite $O(1)$ value in the generic $m\to0$
limit. Second, also $\Psi$ is modified at large distances, and the
modification decays to a different constant. This implies that the
modification is not simply a redefinition of $M_P$. Finally, in FRW
$\tau$ is an ordinary propagating DOF (see footnote \ref{foot}) and,
in contrast to the Minkowski case (see e.g. \cite{Dubovsky:2005dw}),
there is no free time-independent function in the solution. At short
distance both potentials reduce to GR and there is no discontinuity.

With the above explicit solution one can check that in the dS limit
($\ell\to-1$) the potentials reduce to GR, because $\m\to\infty$ and
$\tau\to0$, in agreement with the previous discussion. On the other
hand, also the flat limit $\ell\to 0$ can be safely taken in the last
expression, but the result is not the Minkowski one.  We conclude that the
presence of a curved background removes the linearly growing term at
large distance, or in other words regulates the infrared modification
of the gravitational force.

\medskip
 
One can also find the exact expression for the potential in the phases
where there is no scalar DOF, e.g.\ $m_\eta=0$ (see
section~\ref{m1num}). As we have seen, in this case the kinetic term
of $\tau$ is zero and we we can explicitly solve for its EOM for any
source obtaining an expression similar to (\ref{eq:pot1}). Moreover,
$\tau$ will be of the form
\be
\tau=\frac{ T_{00} }{M_P^2\left(\Delta- M^2\right)}\,,\qquad 
M^2=\frac{q[m_i,a]}{(m_2^2-m_3^3)^2H'}\,,
\ee
where $q[m_i,a]$ is an analytic function of the masses, $a$ and their
derivatives. The correction to Newtonian potential can then be written
in the form
\ba
\label{eq:GImet}
&&\Phi=\Phi_{GR}\left[1+\frac{a^2 m_2^2\, k_1}{(m_2^2-m_3^2)M^2}\left(e^{-M r}-1
+\left[\frac{(m_2^2-3m_3^2+2m_4^2)}{k_1}-1\right]Mr\,e^{-Mr}\right)\right],\nonumber\\
&&\Psi=\Psi_{GR}\left[1+\frac{a^2 \, m_2^2 m_4^2 }{(m_2^2-m_3^2)M^2}\left(e^{-M r}-1\right)
\right] ,
\ea
where 
$$
k_1=(m_2^2-3m_3^2+2m_4^2)+\frac{m_4^2 ( M^2)'}{aH\, M^2}\,.
$$ 
Again we see that the presence of a non trivial background gives rise
to a modification of GR at large distances $r\sim M^{-1}$ (pushed to
infinity for vanishing masses).

At short distances, the potential reduces to GR plus corrections:
\ba
\Phi &=&\Phi_{GR}+
\frac{a^2 m_2^2  T_{00}(m_2^2-3m_3^2+2m_4^2)}{(m_2^2-m_3^2) \Delta ^2
   M_{P}^2}+O\left( \frac{1}{H'\Delta^3}\right).
\ea
On the other hand taking the dS limit carefully we recover
$\Phi=\Phi_{GR}$.

%%%%%%%%%%%%%%%%%%%%%%%%%%%%%%%%%%%%%%%%%%%%%%%%%
\subsection{Coupling to matter for $m_0 =0$}\label{vDVZm0}
%%%%%%%%%%%%%%%%%%%%%%%%%%%%%%%%%%%%%%%%%%%%%%%%%

For the $m_0=0$ case, one can express the potentials in a way similar
to (\ref{eq:pot1}) and (\ref{eq:taup}), this time in terms of $\S$.
The resulting expressions turn out to be very complicated, and here we
will consider explicitly only the ``partially massless'' case
discussed in section~\ref{zerom0}, where no DOF is present. In this
case, we can write the gravitational potentials as
\ba
\label{eq:GIm0}
&&\Phi=\Phi_{GR}+a \left(\frac{(3T_{00}+M_P^2\Delta\S)(a m_4^4+2 m_4^2 H'-2 H (m_4^2)')-2a(3T_{00}+2M_P^2\Delta\S)H^2 m_4^2}{4\Delta^2 H^2 M_P^2}\right),\nonumber\\
&&\Psi=\Psi_{GR}+a^2 m_4^2\left(\frac{3 T_{00}+M_P^2\Delta \S}{2\Delta^2 M_P^2}\right),
\ea
where $\S$ is 
\be
\label{cons:Sig}
 M_P^2\Delta\S=-\frac{3 T_{00}(a m_4^4+2 m_4^2 H'-2 H (m_4^2)'-4aH^2 m_4^2)}{a [m_4^4+2H^2(2m_2^2-3m_4^2)]+2 [m_4^2 H'-H (m_4^2)']}\,.
\ee
The previous two equations indicate that there is no vDVZ discontinuity and
we recover GR in the massless limit.

Some care is needed in the special cases when the numerator or the
denominator of (\ref{cons:Sig}) vanishes. If the denominator vanishes the
theory has an extra gauge invariance (cf.\ section~\ref{zerom0}). As a
result the EMT is coupled consistently only if $T_{00}=0$ unless also the
numerator vanishes. In any case, $\S$ can be set to zero by a gauge
transformation, and the potentials can be read from (\ref{eq:GIm0}).

One can readily see from these expressions that the corrections to the
Newtonian potential simply amount to a linear correction (see
appendix~\ref{app2}), that invalidates the linear approximation at
large distance. This modification vanishes for $m_4=0$, which also
gives $m_1=0$ and $H (3m_3^2-m_2^2)=0$ (cf.\ (\ref{eq:partmass})).

%%%%%%%%%%%%%%%%%%%%%%%%%%%%%%%%%%%%%%%%
\section{Discussion and Conclusions}
\label{sec:conclusions}
%%%%%%%%%%%%%%%%%%%%%%%%%%%%%%%%%%%%%%%%

In this work we have performed a systematic study of Lorentz breaking
massive gravity in a FRW background. For the \emph{\it tensor} and
\emph{\it vector} sectors, the analysis is very close to that of Minkowski
space: both sets of modes acquire independent masses constrained by
phenomenological bounds. For vector modes, the naive strong coupling scale
is similar to that of flat space (see also below).

The most interesting results are in the \emph{scalar} sector where
generically there are two propagating degrees of freedom. For
maximally symmetric spaces, the study of the dispersion relations at
high energy reveals the appearance of \emph{ghost-like} instabilities,
i.e.\ instabilities associated to an infinite volume phase space, that
can be cured only by introducing a momentum space cutoff. Remarkably,
this is not necessarily the case in arbitrary spacetimes: high energy
instabilities can be absent in a FRW background with expanding
horizon, i.e.\ $H'<0$, see (\ref{nog}) and (\ref{eq:generalgrad}).
Indeed, at high energies the sign of the determinant of the mass
matrix ${\cal M}$ (see (\ref{eq:ham}) for the definition) is fixed
in Minkowski and dS whereas in FRW $H'$ enters in the game allowing a
region in the parameter space where ${\cal M}$ is positive definite.
As a drawback, in the limit $H' \to 0$, the theory is strongly coupled
in the scalar sector.

The scalar sector also features a number of phases with less than two
DOF. Generically, in a FRW background the phase $m_0=0$ (which
includes the FP phase) has one scalar DOF. We found the conditions
that make the kinetic term of this mode positive definite,
generalizing the Higuchi bound to LB masses in FRW spaces (see
(\ref{noghostm0}) and (\ref{rootnum})). Moreover, provided that
$m_2^2\geq m_3^2$, high energy instabilities are absent. We also
sketched the method to avoid instabilities at intermediate momenta
(which may be even interesting for cosmological perturbations).

In the presence of curvature there exist situations where the
Lagrangian for the scalar modes becomes particularly simple as
discussed in section~\ref{diffm0}. In particular the invariance under
time diff can be recovered even when $m_1\neq 0$. More interesting is
the case where the absence of scalar DOF is due to a residual gauge
invariance which is absent in flat space (\emph{partially massless
  case}). Taking the FP limit in dS, the condition for having residual
gauge invariance can be solved, and as a result all masses are
determined in terms of the curvature scale (cf.\ (\ref{eq:PFsol})).

Also interesting is the phase $m_1=0$ where in general there is again
a single propagating scalar DOF. For maximally symmetric backgrounds,
the EOM for this scalar do not contain any gradient term and this mode
effectively has no dynamics (zero velocity): it behaves as a
collective mode. For generic FRW with expanding horizon, the
propagating scalar has a dispersion relation that can be made free of
instabilities. 

We have then analyzed how the Newtonian potentials generated by
conserved point-like sources are modified. In the general case they
agree with GR modulo corrections at scales related to the massive
gravity scale. A typical form of the gauge invariant gravitational
potentials is for example (see eqs.\ (\ref{phi}), (\ref{eq:potm1}),
(\ref{eq:GImet}))
\be
\label{eq:res}
\Phi=
\Phi_{GR}\left[1+c_1(e^{-\m_1r}-1)+ c_2\, \m_2r\, e^{-\m_2r}\right],
\ee
where the mass scales $\m_i$ are combinations of curvature and mass
parameters while $c_i$ are dimensionless combinations. This form is
valid also for the $\Psi$ potential. Therefore in the massless limit
or for scales $r\ll (a H)^{-1},\,m^{-1}$ the potentials reduce to the
GR result, which makes these phases potentially very interesting.
Comparing for instance to the $m_1=0$ phase in flat space, where a
linearly growing term invalidates perturbation theory at large
distance \cite{Dubovsky:2004ud}, in curved space the potential
(\ref{eq:res}) is well behaved at large distance without imposing any
fine-tuning in the mass parameters. In this sense, the presence of a
curved background regularizes many of the peculiarities of Minkowski
(also the case $m_\eta=0$, singular in Minkowski, is regular in FRW).
At short distance, the corrections with respect to GR in
(\ref{eq:res}) can be estimated by expanding the exponentials.

In our analysis we found that some of the propagating states can have
small kinetic terms (typically proportional to mass or $H'$) giving
rise to strongly coupled sectors at very low energy. This fact can be
relevant for its possible cosmological implications. For instance
already in the vector sector when $m_1\neq 0$, perturbation theory is
reliable only for $H<\sqrt{m_1 M_P}$. This casts serious doubts on the
possibility to use massive gravity in standard inflation, while
keeping small the LB masses. In fact, to trust the linear
approximation at the standard inflation scale
$H_{\text{inflation}}\simeq 10^{13}\,$GeV one would need
$m_1\gtrsim10^{16}\,$eV, and this would require a severe fine tuning
with respect to the other masses that are constrained by various
gravitational tests (pulsar, solar system tests) to be much smaller
(typically $10^{-21}$ eV). On the other hand, the value of
$H_{\text{inflation}}$ is very model dependent, the only real upper
bound comes from BBN, $T_{RH}\gtrsim 10\,$MeV, i.e.\ $H\gtrsim
10^{-16}\,$GeV. This gives the limit $m_1>10^{-30}\,$eV, well below
any other gravitational constraints on the masses. The cosmological
constraints coming from the analysis of the modified gravitational
perturbation's dynamics are presently under study~\cite{BCNP} .

Let us close with a comment on exact solutions.
Besides the large distance modifications to GR found in this work at
linearized level, some modifications have also been found in exact
(spherically symmetric) solutions of massive
gravity~\cite{Berezhiani:2008nr, babichev}. These solutions exist in
dS space and feature a nonanalytic $r^\gamma$ term in the
gravitational potential.\footnote{As well as a $1/\sqrt{r}$ term in a
  would-be gauge direction (i.e.\ for a Goldstone field). This is
  verified explicitly in bigravity (where
  $m_1=0$)~\cite{Berezhiani:2008nr}, and in a decoupling limit in the
  FP case (Lorentz-invariant
  $m_0=0$)~\cite{babichev}. In~\cite{Bebronne:2009mz}, the solutions
  of ~\cite{Berezhiani:2008nr} have been translated in the goldstone
  formalism and extended numerically to other nonlinear lagrangians.}
Thus, they differ also asymptotically from the linearized
gravitational potentials found in this work, which may be understood
from the presence of long-range instantaneous interactions at
linearized level. Therefore also for many of the phases analyzed in
the present work, one may expect important non-linear effects even at
large distances.

Finally, our study suggests that the analysis of perturbations around
other nontrivial backgrounds may also unveil phases where the
perturbations have a stable spectrum, and is thus of definite
interest. Of main importance would be a dedicated study addressing
perturbations and their stability in the (exact) gravitational
background produced by a star.

\acknowledgments

This work was partially supported by the EU FP6 Marie Curie Research
\& Training Network "UniverseNet" (MRTN-CT-2006-035863)". The work of
D.~B. is founded by the Swiss Science Foundation. D.~B. would like to
thank the warm hospitality of the LNGS, where this work was initiated.
L.~P. would like to thank INFN and LNGS for the support in the
aftermath of earthquake that struck L'Aquila.  The comments from
participants of the SW3 conference are also acknowledged.

\begin{appendix}

%%%%%%%%%%%%%%%%%%%%%%%%%%%%%%%%%%%%%%%%%%%%%%%%
\section{The phase $m_0=0$ in the generic case}\label{app1}
%%%%%%%%%%%%%%%%%%%%%%%%%%%%%%%%%%%%%%%%%%%%%%%%%

In this appendix we present the explicit analysis of the dynamical
degrees of freedom of the case $m_0=0$ discussed in
section~\ref{sec:m0g}. The EOM for $\t$ yield the constraint
\ba
\tau &=& \frac{a^2}{D}\bigg\{
\Sigma'\Big[4  \Delta  \left(m^2_1-2 m^2_4\right)H \Big]
+\Sigma \Big[2\left(4 \Delta -3 a^2
   m^2_1\right)\left(
 a m_\mu^2   H^2 -  (m^2_4)' H
+m^2_4H'\right)
\nb\\
&&\qquad\qquad\qquad\qquad\qquad\qquad\qquad{}
+am^2_4  m^2_1 \left(2 \Delta -3 a^2 m^2_4\right)\Big]\bigg\}\,,
\label{eq:tau}
\\
D&=&4 \Delta ^2\Big[a m^2_1+4 H'\Big]
   -12 \Delta \Big[2 a m_\mu^2  H^2-2 (m_4^2)' H+a m^2_1 m^2_4+\left(m^2_1+2
                  m^2_4\right) H'\Big] a^2
\nb\\
&&{}
+9 m^2_1 \Big[2 a m_\mu^2  H^2 +a (m^2_4)^2+2(m_4^2 H'- (m^2_4)' H)\Big]
   a^4\,,\nb
\ea
where $m_\mu^2=3(m_3^2-m_4^2)-m_2^2$.  Once $\tau$ is substituted in
the action, we get a (quite involved) effective action for
$\Sigma$ whose kinetic part is written in (\ref{kinm0}).
%
%\be\label{kinm0}
% \frac{\cal K}{ a^4M_P^2} = \frac{3a^3m_1^2
%\left[a (m_4^2)^2 + 2 (a m_\mu^2H^2-H (m_4^2)'+m_4^2 H') \right]
%-4[am_4^2 (m_1^2-m_4^2)+m_1^2H']\Delta}
%{a m_1^2(2\Delta-3a^2m_4^2)^2-2(4\Delta-3a^2 m_1^2)[3 a^2H( aH m_\mu^2- (m_4^2)')
%-\left(2 \Delta
%   -3 a^2 m^2_4\right) H']} ,
%\ee
%
Requiring the positivity of the kinetic energy (no ghost) for large momenta
we find  the condition (\ref{noghostm0}). As we saw, the kinetic
term is always positive at small momenta.
%
%\be
%\label{noghostm0:ap}
%\frac{a  m_4^2 (m_1^2- m_4^2)+m_1^2  H'}{ a m_1^2+4 H'} >0 \, .
%\ee
%
%On the other hand at small momenta the kinetic term 
%reduces to 
% $${\mathcal K}|_{\Delta=0}= a^2 M_P^2/3,$$
%which is, remarkably,
%always positive. 

To understand when $\K$ is positive also at intermediate momenta,
first notice that $\K$ is expressed as a fraction of two polynomials
with different roots\footnote{The two roots coincide only when
\be
m_1^2(m_1^2-2m_4^2)^2[a(m_4^4+2 H^2 m_\m^2)+2m_4^2 H'-H (m_4^2)']=0.
\ee
If the second factor cancels, we find a very simple kinetic term which
is always positive. When the last factor cancels, the kinetic term is
positive at any scale provided that (\ref{noghostm0}) holds. Finally,
when both terms cancel, the constraint (\ref{eq:tau}) reduces to
$\tau=0$ and the whole action is much simpler.} in the variable
$\Delta$. For the fraction to keep its sign those roots must be either
at $\Delta >0$ or be absent. The numerator is a linear polynomial, and
setting its root at positive $\Delta$ corresponds to the condition
(\ref{rootnum}). The denominator is a second order polynomial that,
once the coefficient of $\Delta^2$ is taken as a common factor, has a
the zeroth order term
\be
\label{rootnum2}
9a^4m_1^2\left(\frac{a(m_4^4+2 H^2 m_\m^2)+2 m_4^2 H'-2 H (m_4^2)'}{a m_1^2+4 H'}\right),
\ee
which, from (\ref{noghostm0}) and (\ref{rootnum}) is positive. Thus,
the product of the roots of the polynomial is positive. Besides, the
term proportional to $\Delta$ in the denominator reads
\be
-12\left(\frac{a(m_4^4+2 H^2 m_\m^2)+2 m_4^2 H'-2 H (m_4^2)'}{9a^2(a m_1^2+4 H')}\right)-12a^2\left(
\frac{a  m_4^2 (m_1^2- m_4^2)+m_1^2  H'}{ a m_1^2+4 H'}\right),
\ee
which from (\ref{noghostm0}) and (\ref{rootnum}) is negative definite
for $m_1^2\geq 0$. This finally means that in the case $m_1^2\geq 0$
(required for stability of the vector sector), both roots are positive
and instabilities in the kinetic term are absent at any scale provided
that the inequalities (\ref{noghostm0}) and (\ref{rootnum}) are
satisfied.

\medskip

Concerning the potential term, the analysis is more involved. The
expansion in $\Delta$ provides a useful tool to analyze the absence of
instabilities at any momentum scale. The potential term $\M$ can be
written as (\ref{eq:m0pot}).
The absence of gradient instabilities in the ultraviolet and infrared regimes, equivalent to the positivity
of ${\cal M}$ in these regimes, requires 
\be
%{\cal M} > 0 \Rightarrow
\begin{cases}
\text{at large momenta} & m_2^2>m_3^2 \\
\text{at small momenta} & m_2^2 > 0 \,.
\end{cases}
\ee
At intermediate scales, the instabilities are {\em Jeans-like}.
The  potential ${\mathcal L}_V$ is
quartic in $\Delta$ which does not allow to find its zero exactly. Nevertheless, 
imposing that it is free from instabilities at high energy
scales and at zero momentum we know that it will be positive definite at any scale provided
that its minima in the interval $\Delta\in (-\infty,0]$ are below
  zero. These minima can be exactly localized as they corresponds to the
  solutions of
\be
c+2d\Delta+3e\Delta^2+4(m_2^2-m_3^2)\Delta^3=0.
\ee
{}From the fact that we have at much two minima localized in the
interval $\Delta\in (-\infty,0]$, and yet some extra freedom in the
choice of the mass functions, we expect to find a large class of
lagrangians with a well defined potential (see \cite{Blas:2008uz}).

%%%%%%%%%%%%%%%%%%%%%%%%%%%%%%%%%%%%%%%%%%%%%%%%%%%%%%%%%%%%%%%%%
\section{Gravitational Green's Functions}\label{app2}
%%%%%%%%%%%%%%%%%%%%%%%%%%%%%%%%%%%%%%%%%%%%%%%%%%%%%%%%%%%%%%%%%

Once the  Newtonian potentials are worked out in momentum space, they can 
easily be Fourier-transformed to the physical position $r$-space ($r=|\vec{x}|$).
The potentials $\Phi$ and $\Psi$ found in this work are always of the kind
\ba
\label{pole}
\Phi=\frac{\rm Polynomial(\Delta^n+\cdots)}{\rm Polynomial(\Delta^{n+2}+\cdots)}=
\sum_i\frac{Z_{i}}{(\Delta-M_{i}^2)^i}
\ea
where $Z_{i}$ and $M_{i}$ are functions of the background and mass
parameters. Once the fraction has been decomposed in poles, we can use
the following correspondence to directly read the $r$ dependence:
\ba
\frac{1}{\Delta}\rightarrow \frac{1}{r},\qquad
\frac{1}{\Delta^2}\rightarrow r,\qquad \frac{1}{\Delta-m^2}
\rightarrow \frac{e^{-mr}}{r},\qquad \frac{1}{(\Delta-m^2)^2}
\rightarrow \frac{e^{-mr}}{m},\quad {\rm etc},
\ea
for suitable choices of integration constants. 

\end{appendix}

%%%%%%%%%%%%%%%%%%%%%%%%%%%%%%%%%%%%%%%%%%%%%%%%%%%%%%%%%%%%%%
%%%%%%%%%%%%%%%%%%%%%%%%%%%%%%%%%%%%%%%%%%%%%%%%%%%%%%%%%%%%%%


\begin{thebibliography}{99}
%%%%%%%%%%%%%%%%%%%%%%%%%%%%%%%%%%%%%%%%%%%%%%%%%%%%%%%%%%%%%%
%%%%%%%%%%%%%%%%%%%%%%%%%%%%%%%%%%%%%%%%%%%%%%%%%%%%%%%%%%%%%%

  %\cite{Rubakov:2008nh}
\bibitem{Rubakov:2008nh}
  V.~A.~Rubakov and P.~G.~Tinyakov,
  %``Infrared-modified gravities and massive gravitons,''
  Phys.\ Usp.\  {\bf 51} (2008) 759.
%  [arXiv:0802.4379 [hep-th]].
  %%CITATION = PHUSE,51,759;%%


%\cite{Blas:2008uz}
\bibitem{Blas:2008uz}
  D.~Blas,
  %``Aspects of Infrared Modifications of Gravity,''
  {\tt arXiv:0809.3744 [hep-th]}.
  %%CITATION = ARXIV:0809.3744;%%


\bibitem{vDVZ}
%\cite{vanDam:1970vg}
%\bibitem{vanDam:1970vg}
H.~van Dam and M.~J.~Veltman,
%``Massive And Massless Yang-Mills And Gravitational Fields,''
Nucl.\ Phys.\ B {\bf 22} (1970) 397.
%%CITATION = NUPHA,B22,397;%%
%\bibitem{ZAK}
V.I.Zakharov, JETP Lett {\bf 12}, 312 (1970).
%\cite{Iwasaki:uz}
%\bibitem{Iwasaki:uz}
Y.~Iwasaki,
%``Consistency Condition For Propagators,''
Phys.\ Rev.\ D {\bf 2} (1970) 2255.
%%CITATION = PHRVA,D2,2255;%%


%\cite{ArkaniHamed:2002sp}
\bibitem{ArkaniHamed:2002sp}
  N.~Arkani-Hamed, H.~Georgi and M.~D.~Schwartz,
  %``Effective field theory for massive gravitons and gravity in theory space,''
  Annals Phys.\  {\bf 305} (2003) 96.
%  [arXiv:hep-th/0210184].
  %%CITATION = APNYA,305,96;%%

%\cite{Deffayet:2001uk}
\bibitem{Deffayet:2001uk}
  C.~Deffayet, G.~R.~Dvali, G.~Gabadadze and A.~I.~Vainshtein,
  %``Nonperturbative continuity in graviton mass versus perturbative
  %discontinuity,''
  Phys.\ Rev.\ D {\bf 65} (2002) 044026.
%  [arXiv:hep-th/0106001].
  %%CITATION = HEP-TH 0106001;%%


%\cite{Dvali:2006su}
\bibitem{Dvali:2006su}
  G.~Dvali,
  %``Predictive Power of Strong Coupling in Theories with Large Distance
  %Modified Gravity,''
  New J.\ Phys.\  {\bf 8} (2006) 326.
%  [arXiv:hep-th/0610013].
  %%CITATION = NJOPF,8,326;%%


%\cite{Rubakov:2004eb}
\bibitem{Rubakov:2004eb}
  V.~A.~Rubakov,
  %``Lorentz-violating graviton masses: Getting around ghosts, low strong
  %coupling scale and VDVZ discontinuity,''
  {\tt arXiv:\hepth{0407104}}.
  %%CITATION = HEP-TH/0407104;%%



%\cite{Dubovsky:2004sg}
\bibitem{Dubovsky:2004sg}
  S.~L.~Dubovsky,
  %``Phases of massive gravity,''
  JHEP {\bf 0410} (2004) 076.
%  [arXiv:hep-th/0409124].
  %%CITATION = JHEPA,0410,076;%%



  


\bibitem{vDVZAdS}
%\cite{Porrati:2000cp}
%\bibitem{Porrati:2000cp}
  M.~Porrati,
  %``No van Dam-Veltman-Zakharov discontinuity in AdS space,''
  Phys.\ Lett.\  B {\bf 498} (2001) 92.
%  [arXiv:hep-th/0011152].
  %%CITATION = PHLTA,B498,92;%%
  
%\cite{Karch:2001jb}
%\bibitem{Karch:2001jb}
  A.~Karch, E.~Katz and L.~Randall,
  %``Absence of a VVDZ discontinuity in AdS(AdS),''
  JHEP {\bf 0112} (2001) 016.
%  [arXiv:hep-th/0106261].
  %%CITATION = JHEPA,0112,016;%%
  
%\cite{Kogan:2000uy}
%\bibitem{Kogan:2000uy}
  I.~I.~Kogan, S.~Mouslopoulos and A.~Papazoglou,
  %``The m --> 0 limit for massive graviton in dS(4) and AdS(4): How to
  %circumvent the van Dam-Veltman-Zakharov discontinuity,''
  Phys.\ Lett.\  B {\bf 503} (2001) 173.
%  [arXiv:hep-th/0011138].
  %%CITATION = PHLTA,B503,173;%%

%\cite{Gabadadze:2008ha}
%\bibitem{Gabadadze:2008ha}
  G.~Gabadadze, A.~Iglesias and Y.~Shang,
  %``General Massive Spin-2 on de Sitter Background,''
  {\tt arXiv:0809.2996 [hep-th]}.
  %%CITATION = ARXIV:0809.2996;%%


%\cite{Creminelli:2005qk}
\bibitem{Creminelli:2005qk}
  P.~Creminelli, A.~Nicolis, M.~Papucci and E.~Trincherini,
  %``Ghosts in massive gravity,''
  JHEP {\bf 0509} (2005) 003.
%  [arXiv:hep-th/0505147].
  %%CITATION = JHEPA,0509,003;%%

 %\cite{Nair:2008yh}
\bibitem{Nair:2008yh}
  V.~P.~Nair, S.~Randjbar-Daemi and V.~Rubakov,
  %``Massive Spin-2 fields of Geometric Origin in Curved Spacetimes,''
  {\tt arXiv:0811.3781 [hep-th]}.
  %%CITATION = ARXIV:0811.3781;%%


%\cite{Horava:2009uw}
\bibitem{Horava:2009uw}
  P.~Horava,
  %``Quantum Gravity at a Lifshitz Point,''
  {\tt arXiv:0901.3775 [hep-th]}.
  %%CITATION = ARXIV:0901.3775;%%





%\cite{Dubovsky:2004ud}
\bibitem{Dubovsky:2004ud}
  S.~L.~Dubovsky, P.~G.~Tinyakov and I.~I.~Tkachev,
  %``Massive graviton as a testable cold dark matter candidate,''
  Phys.\ Rev.\ Lett.\  {\bf 94} (2005) 181102.
%  [arXiv:hep-th/0411158].
  %%CITATION = PRLTA,94,181102;%%

%\cite{ArkaniHamed:2003uy}
\bibitem{ArkaniHamed:2003uy}
  N.~Arkani-Hamed, H.~C.~Cheng, M.~A.~Luty and S.~Mukohyama,
  %``Ghost condensation and a consistent infrared modification of gravity,''
  JHEP {\bf 0405} (2004) 074.
%  [arXiv:hep-th/0312099].
  %%CITATION = JHEPA,0405,074;%%



%\cite{Bebronne:2007qh}
\bibitem{Bebronne:2007qh}
  M.~V.~Bebronne and P.~G.~Tinyakov,
  %``Massive gravity and structure formation,''
  Phys.\ Rev.\  D {\bf 76} (2007) 084011.
%  [arXiv:0705.1301 [astro-ph]].
  %%CITATION = PHRVA,D76,084011;%%


%\cite{Damour:2002ws}
\bibitem{Damour:2002ws}
  T.~Damour and I.~I.~Kogan,
  %``Effective Lagrangians and universality classes of nonlinear bigravity,''
  Phys.\ Rev.\ D {\bf 66} (2002) 104024.
%  [arXiv:hep-th/0206042].
  %%CITATION = HEP-TH 0206042;%%


%\cite{Blas:2007zz}
\bibitem{Blas:2007zz}
  D.~Blas, C.~Deffayet and J.~Garriga,
  %``Bigravity and Lorentz-violating Massive Gravity,''
  Phys.\ Rev.\  D {\bf 76} (2007) 104036.
%  [arXiv:0705.1982 [hep-th]].
  %%CITATION = PHRVA,D76,104036;%%



%\cite{Berezhiani:2007zf}
\bibitem{Berezhiani:2007zf}
  Z.~Berezhiani, D.~Comelli, F.~Nesti and L.~Pilo,
  %``Spontaneous Lorentz breaking and massive gravity,''
   Phys.\ Rev.\ Lett.\  {\bf 99} (2007) 131101. 
% [arXiv:hep-th/0703264].
  %%CITATION = HEP-TH/0703264;%%


\bibitem{Berezhiani:2009kx}

  N.~Rossi,
  %``Dark Halo or Bigravity?,''
  Eur.\ Phys.\ J.\ ST {\bf 163} (2008) 291
  {\tt [arXiv:0902.0072 [astro-ph.CO]]}.
  %%CITATION = 00619,163,291;%%


  Z.~Berezhiani, L.~Pilo and N.~Rossi,
  %``Mirror Matter, Mirror Gravity and Galactic Rotational Curves,''
  {\tt arXiv:0902.0146 [astro-ph.CO]}.
   %%CITATION = ARXIV:0902.0146;%%

  Z.~Berezhiani, F.~Nesti, L.~Pilo and N.~Rossi,
  %``Gravity Modification with Yukawa-type Potential: Dark Matter and Mirror
  %Gravity,''
  {\tt arXiv:0902.0144 [hep-th]}.
  %%CITATION = ARXIV:0902.0144;%%

M.~Banados, A.~Gomberoff, D.~C.~Rodrigues and C.~Skordis,
  %``A note on bigravity and dark matter,''
  Phys.\ Rev.\  D {\bf 79} (2009) 063515.
  %[arXiv:0811.1270 [gr-qc]].
%%CITATION = PHRVA,D79,063515;%%
%\cite{Berezhiani:2008nr}
\bibitem{Berezhiani:2008nr}
  Z.~Berezhiani, D.~Comelli, F.~Nesti and L.~Pilo,
  %``Exact Spherically Symmetric Solutions in Massive Gravity,''
  JHEP {\bf 0807} (2008) 130.
  %  [arXiv:0803.1687 [hep-th]].
  %%CITATION = JHEPA,0807,130;%%
  
%\cite{Deffayet:2008zz}
\bibitem{Deffayet:2008zz}
  C.~Deffayet,
  %``Spherically symmetric solutions of massive gravity,''
  Class.\ Quant.\ Grav.\  {\bf 25} (2008) 154007.
  %%CITATION = CQGRD,25,154007;%%



%\cite{Koroteev:2009xd}
\bibitem{Koroteev:2009xd}
  P.~Koroteev and M.~Libanov,
  %``Spectra of Field Fluctuations in Braneworld Models with Broken Bulk Lorentz
  %Invariance,''
  Phys.\ Rev.\  D {\bf 79} (2009) 045023
  {\tt [arXiv:0901.4347 [hep-th]]}.
  %%CITATION = PHRVA,D79,045023;%%

%\cite{Libanov:2005vu}
\bibitem{Libanov:2005vu}
  M.~V.~Libanov and V.~A.~Rubakov,
  %``More about spontaneous Lorentz-violation and infrared modification of
  %gravity,''
  JHEP {\bf 0508} (2005) 001
  {\tt [arXiv:\hepth{0505231}]}.
  %%CITATION = JHEPA,0508,001;%%

%\cite{Dubovsky:2005dw}
\bibitem{Dubovsky:2005dw}
  S.~L.~Dubovsky, P.~G.~Tinyakov and I.~I.~Tkachev,
  %``Cosmological attractors in massive gravity,''
  Phys.\ Rev.\  D {\bf 72} (2005) 084011.
%  [arXiv:hep-th/0504067].
  %%CITATION = PHRVA,D72,084011;%%

%\cite{Grisa:2008um}
\bibitem{Grisa:2008um}
  L.~Grisa,
  %``Lorentz-Violating Massive Gravity in Curved Space,''
  JHEP {\bf 0811} (2008) 023.
%  [arXiv:0803.1137 [hep-th]].
  %%CITATION = JHEPA,0811,023;%%

%\cite{Mukohyama:2006be}
%\bibitem{Mukohyama:2006be}
  S.~Mukohyama,
  %``Accelerating universe and cosmological perturbation in the ghost
  %condensate,''
  JCAP {\bf 0610} (2006) 011.
%  [arXiv:hep-th/0607181].
  %%CITATION = JCAPA,0610,011;%%

%\bibitem{Damour:2002wu}
  T.~Damour, I.~I.~Kogan and A.~Papazoglou,
  %``Non-linear bigravity and cosmic acceleration,''
  Phys.\ Rev.\  D {\bf 66} (2002) 104025.
%  [arXiv:hep-th/0206044].
  %%CITATION = PHRVA,D66,104025;%%

\bibitem{BCNP}
  D.~Blas, D.~Comelli, F.~Nesti and L.~Pilo,
  {\em in preparation}.
  %

%\cite{Ford:1980up}
\bibitem{Ford:1980up}
  L.~H.~Ford and H.~Van Dam,
  %``The Impossibility Of A Nonzero Rest Mass For The Graviton,''
  Nucl.\ Phys.\  B {\bf 169} (1980) 126.
  %%CITATION = NUPHA,B169,126;%%

%\cite{Cline:2003gs}
\bibitem{Cline:2003gs}
  J.~M.~Cline, S.~Jeon and G.~D.~Moore,
  %``The phantom menaced: Constraints on low-energy effective ghosts,''
  Phys.\ Rev.\  D {\bf 70} (2004) 043543.
%  [arXiv:hep-ph/0311312].
  %%CITATION = PHRVA,D70,043543;%%

%\cite{Dubovsky:2005xd}
\bibitem{Dubovsky:2005xd}
  S.~Dubovsky, T.~Gregoire, A.~Nicolis and R.~Rattazzi,
  %``Null energy condition and superluminal propagation,''
  JHEP {\bf 0603} (2006) 025.
  %[arXiv:hep-th/0512260].
  %%CITATION = JHEPA,0603,025;%%
  
%\cite{Libanov:2008mk}
\bibitem{Libanov:2008mk}
  M.~V.~Libanov, V.~A.~Rubakov, O.~S.~Sazhina and M.~V.~Sazhin,
  %``CMB anisotropy induced by tachyonic perturbations of dark energy,''
  J.\ Exp.\ Theor.\ Phys.\  {\bf 108} (2009) 226
  {\tt [arXiv:0812.1459 [astro-ph]]}.
  %%CITATION = JTPHE,108,226;%%

%\cite{Abbott:1981ff}
\bibitem{Abbott:1981ff}
  L.~F.~Abbott and S.~Deser,
  %``Stability Of Gravity With A Cosmological Constant,''
  Nucl.\ Phys.\  B {\bf 195} (1982) 76.
  %%CITATION = NUPHA,B195,76;%%

%\cite{Deser:2001wx}
\bibitem{Deser:2001wx}
  S.~Deser and A.~Waldron,
  %``Stability of massive cosmological gravitons,''
  Phys.\ Lett.\  B {\bf 508} (2001) 347.
%  [arXiv:hep-th/0103255].
  %%CITATION = PHLTA,B508,347;%%
  
%\cite{Polchinski:1992ed}
\bibitem{Polchinski:1992ed}
  J.~Polchinski,
  %``Effective Field Theory And The Fermi Surface,''
  TASI 92, {\tt arXiv:\hepth{9210046}}.
  %%CITATION = HEP-TH/9210046;%%

%\cite{Mukhanov:1990me}
\bibitem{Mukhanov:1990me}
  V.~F.~Mukhanov, H.~A.~Feldman and R.~H.~Brandenberger,
  %``Theory of cosmological perturbations. Part 1. Classical perturbations. Part
  %2. Quantum theory of perturbations. Part 3. Extensions,''
  Phys.\ Rept.\  {\bf 215} (1992) 203.
  %%CITATION = PRPLC,215,203;%%

%\cite{Dvali:2008em}
\bibitem{Dvali:2008em}
  G.~Dvali, O.~Pujolas and M.~Redi,
  %``Non Pauli-Fierz Massive Gravitons,''
  Phys.\ Rev.\ Lett.\  {\bf 101} (2008) 171303.
  %[arXiv:0806.3762 [hep-th]].
  %%CITATION = PRLTA,101,171303;%%

%\cite{Fierz:1939ix}
\bibitem{Fierz:1939ix}
  M.~Fierz and W.~Pauli,
  %``On relativistic wave equations for particles of arbitrary spin in an
  %electromagnetic field,''
  Proc.\ Roy.\ Soc.\ Lond.\  A {\bf 173} (1939) 211.
  %%CITATION = PRSLA,A173,211;%%

%\cite{Higuchi:1986py}
\bibitem{Higuchi:1986py}
  A.~Higuchi,
  %``FORBIDDEN MASS RANGE FOR SPIN-2 FIELD THEORY IN DE SITTER SPACE-TIME,''
  Nucl.\ Phys.\  B {\bf 282} (1987) 397.
  %%CITATION = NUPHA,B282,397;%%
  

%\cite{Deser:1983mm}
\bibitem{Deser:1983mm}
  S.~Deser and R.~I.~Nepomechie,
  %``Gauge Invariance Versus Masslessness In De Sitter Space,''
  Annals Phys.\  {\bf 154} (1984) 396.
  %%CITATION = APNYA,154,396;%%

%\cite{Arun:2009pq}
\bibitem{Arun:2009pq}
  K.~G.~Arun and C.~M.~Will,
  %``Bounding the mass of the graviton with gravitational waves: Effect of
  %higher harmonics in gravitational waveform templates,''
  {\tt arXiv:0904.1190 [gr-qc]}.
  %%CITATION = ARXIV:0904.1190;%%
  
 %\cite{Pshirkov:2008nr}
%\bibitem{Pshirkov:2008nr}
  M.~Pshirkov, A.~Tuntsov and K.~A.~Postnov,
  %``Constraints on the massive graviton dark matter from pulsar timing and
  %precision astrometry,''
  Phys.\ Rev.\ Lett.\  {\bf 101} (2008) 261101.
%  [arXiv:0805.1519 [astro-ph]].
  %%CITATION = PRLTA,101,261101;%%


\bibitem{Bessada:2009qw}
  D.~Bessada and O.~D.~Miranda,
  %``CMB Polarization and Theories of Gravitation with Massive Gravitons,''
  Class.\ Quant.\ Grav.\  {\bf 26} (2009) 045005.
%  [arXiv:0901.1119 [gr-qc]].
  %%CITATION = CQGRD,26,045005;%%


\bibitem{babichev}
E.~Babichev, C.~Deffayet and R.~Ziour,
  %``The Vainshtein mechanism in the Decoupling Limit of massive gravity,''
   {\tt arXiv:arXiv:0901.0393 [hep-th]}.
%%CITATION = ARXIV:0901.0393;%%

\bibitem{Bebronne:2009mz}
  M.~V.~Bebronne and P.~G.~Tinyakov,
  %``Black hole solutions in massive gravity,''
   {\tt arXiv:0902.3899 [gr-qc]}.
%\bibitem{Babichev:2009us}

% %\cite{Will:2005va}
% \bibitem{Will:2005va}
%   C.M.~Will,
%   %``The confrontation between general relativity and experiment,''
%   {\tt arXiv:gr-qc/0510072}.
%   %%CITATION = GR-QC/0510072;%%


\end{thebibliography}
\end{document}